\documentclass[twocolumn,showpacs,amsmath,amssymb,showkeys,floatfix,prl]{revtex4}
\usepackage{amsfonts,amsmath}
\usepackage{graphicx}
\usepackage{dcolumn}
\usepackage{bm, bbm}

\begin{document}

\title{Parametric enhancement of flavor oscillation in a three-neutrino framework}

\author{Kara M.~Merfeld}
\author{David C.~Latimer}
\affiliation{Department of Physics, University of Puget Sound,
Tacoma, WA 98416-1031
}
\date{\today}

\begin{abstract}
When neutrinos travel through matter with a periodic density profile, the neutrino oscillation probability can be enhanced if certain conditions are satisfied.  In a two-neutrino framework, the condition for parametric resonance is known.  Herein, we consider the analogous parametric resonance condition within the context of a full three-neutrino framework with two oscillation scales.  For energies in the range of hundreds of MeV to a few GeV, we find that neutrino oscillation can be parametrically enhanced if two approximate relations are satisfied.  The first is similar to the two-neutrino parametric resonance condition while the second involves the other oscillation scale.  Treating the Earth's density as piecewise constant, we show that oscillations in this energy range can be enhanced between two- and threefold.
\end{abstract}

\maketitle
\section{Introduction}

Now that the existence of neutrino mass is firmly established,   experimentalists are tasked with the job of improving the precision of our knowledge of the parameters which characterize three-flavor neutrino oscillations:  three mixing angles, one Dirac CP phase, and two mass-squared differences.   Though the CP phase is ill constrained and the ordering of the mass eigenstates is unknown, global analyses indicate that the three mixing angles and mass-squared differences are known to a precision on the order of a few percent \cite{Capozzi:2013csa,Forero:2014bxa}.  Before reaching such a level of precision, a single oscillation experiment could be reasonably understood within the context of two neutrinos, but the improved precision of accelerator neutrino experiments \cite{Adamson:2014vgd, Abe:2014ugx} requires one to consider terms beyond the effective two-neutrino approximation. 

Since the full three-neutrino spectrum must be considered in current analyses of oscillation experiments, we wish to examine the phenomenon of  parametric enhancement of  flavor oscillation in a three-neutrino framework. Though matter is largely transparent to neutrinos, the oscillation parameters for neutrinos traveling through matter are effectively modified in an energy and density dependent way  \cite{ms,w}.  If the matter through which neutrinos travel has the appropriate periodic density profile, then the flavor oscillation probability can be parametrically enhanced \cite{ermilova,akhmedov_1}.  This phenomenon of parametrically enhanced neutrino oscillations has an analog in mechanical systems.
 For mechanical oscillators, the amplitude of oscillation can be enhanced if the oscillation parameters change at roughly twice the natural frequency of the oscillator.  
As an example, a pendulum whose support oscillates vertically at twice the pendulum's natural frequency will increase in amplitude no matter how small the initial amplitude  \cite{ll_mechanics}. 

Since its initial discovery, parametric enhancement of neutrino oscillations has been extensively studied through both analytical and numerical means \cite{ermilova,akhmedov_1,krastev,liu_smirn, liu_mikh_smirn,petcov_param,akh_long,akh_atmos_SK,floquet,chizhov,kimura,param_highE, akh_13, akh_12, akh_nu_osc_cpv,param_fourier, hay}. In Refs.~\cite{liu_smirn, liu_mikh_smirn}, it was determined that the Earth's interior might provide a suitable matter density profile for which to realize parametric resonance.  As a first approximation, the Earth's density can be divided into two regions:  a high density core surrounded by a lower density mantle \cite{prem}.  Neutrino trajectories which pass through the core sample one and one-half periods of a periodic matter profile.  Despite traveling through fewer than two full periods, the Earth's density profile can parametrically enhance the oscillation probability of atmospheric neutrinos  \cite{akh_atmos_SK,param_highE,akh_13, akh_12,akh_nu_osc_cpv,param_fourier, hay}.

A periodic potential consisting of two piecewise constant regions of differing densities is often referred to as a ``castle-wall" potential because, when plotted, the function resembles these walls' crenellation.  In a two-neutrino scheme, exact analytic solutions through such castle-wall profiles exist, and these results serve as a fundamental tool for understanding parametric enhancement for core-crossing trajectories
\cite{akhmedov_1,akh_long,floquet}.    Relatively exhaustive semi-analytic and numerical studies for neutrino oscillations in the earth were done in Refs.~\cite{akh_13, akh_12} where resonance regions are shown to follow from generalized amplitude and phase conditions.  
To consider analytically parametric resonance in a three-neutrino framework, one typically introduces relevant approximations as a means to reduce the problem to an effective two-neutrino system.  In this manner, one may incorporate the Dirac CP phase into the analysis, something not possible in a pure two-neutrino theory.  Using these techniques, the authors of Refs.~\cite{akh_12,akh_nu_osc_cpv} study the consequences of CP violation to the oscillation probability for multi-GeV neutrinos traveling through the earth; particular attention is paid to the interference between oscillations due to the $\Delta_{21}$ and $\Delta_{31}$ mass-squared differences. 
In Ref.~\cite{hay}, the authors use approximations relevant for sub-GeV atmospheric neutrinos to study the impact of CP violation on the parametric resonance condition for neutrinos traveling through the Earth.

The exact solution for the parametric resonance condition in the two-neutrino framework with the castle-wall potential is tractable because simple expressions exist for the single effective mixing angle and mass-squared difference in matter.  This makes it rather easy to determine the time evolution of a neutrino state through a region of constant density. Using  well known identities for Pauli matrices, the time evolution operator through constant density matter can simply expressed, and then, the time-evolution operator through one period of a castle-wall profile can be expressed in a compact analytical form.  In a three-neutrino framework, this is not the case.   Expressions do exist for the effective two independent mass-squared differences and three mixing angles in matter of constant density, but they are rather opaque.  Furthermore, compact expressions for the time evolution of a neutrino state through matter of constant density do not exist.  

In order to study three-neutrino parametric resonance, we will develop a relatively simple expression for the time evolution operator relevant for neutrinos traveling through matter of constant density.
By making our matter Hamiltonian traceless, we can express it in terms of the Gell-Mann matrices, and then through exponetiation, we are able to write the time-evolution operator as a linear combination of the identity and the Gell-Mann matrices with coefficients given by elements of the Hamiltonian.  We will then use this formulation to consider the propagation of neutrinos through a varying density profile in pursuit of the parametric resonance condition within a full three-neutrino framework.  For a castle-wall profile, we will first consider active mixing amongst two of the neutrinos, effectively recovering the two neutrino results from Refs.~\cite{akhmedov_1,akh_long,floquet}.  Then, we continue to consider the more general three neutrino picture.  We conclude with an application of the results to neutrinos which traverse the Earth's core.

\section{Oscillation in matter of constant density}

Neutrinos produced in weak interactions have definite flavor:  electron, muon, or tau; yet these flavor states are superpositions of states of definite mass $m_j$.  If one such mass state $\nu_j$ is an energy eigenstate, then it evolves, in vacuum, according to 
\begin{equation}
i \partial_t \nu_j = E_j \nu_j \,,
\end{equation}
where we employ natural units, $c = \hbar =1$. Since neutrinos are ultrarelativistic, we approximate the energy as $E_j \approx p+ m_j^2/2E$.  The flavor states $\nu_\sigma$ (with $\sigma = e, \mu, \tau$) are related to the mass eigenstates via a unitary mixing matrix $U$, $\nu_\sigma = U_{\sigma j} \nu_j$ (summation implied).  A column vector, $\nu$, representing the flavor states then evolves, in vacuum, according to
\begin{equation}
i \partial_t \nu = \frac{1}{2E} U \mathcal{M} U^\dagger \nu \, ,  \label{vac_ode}
\end{equation}
where we define the matrix $\mathcal{M} = 
\mathrm {diag} (m_1^2, m_2^2, m_3^2)$.  To simplify notation, we have subtracted from the Hamiltonian a multiple of the identity, $p\mathbbm{1}$.  This common momentum results in an overall unmeasurable phase, so we omit it.  

Defining the vacuum Hamiltonian $H_0 :=  \frac{1}{2E} U \mathcal{M} U^\dagger$, the time evolution of a flavor state is $\nu(t) = \mathcal{U}(t)  \nu(0)$, where the time evolution operator is given by $\mathcal{U}(t) := \exp[-iH_0t]$.  An explicit expression for the time evolution operator can be easily achieved by rotating the Hamiltonian to the mass basis
\begin{equation}
\mathcal{U}(t) = U \left( \begin{array}{ccc}
e^{-im_1^2 L/2E} & 0 &0\\
0& e^{-im_2^2 L/2E}&0\\
0&0& e^{-im_3^2 L/2E}
\end{array} \right) U^\dagger . \label{time_evol_vac}
\end{equation}
With this expression, we can compute the neutrino flavor oscillation probability.  

Supposing that a source produces neutrinos of $\sigma$-flavor, $\nu(0)=\nu_\sigma$, the probability that they are detected as $\rho$-flavor at a time $t$ is given by 
\begin{equation}
\mathcal{P}_{\sigma \rho}(t) = |\langle \nu_\rho | \nu(t) \rangle|^2 = |[\mathcal{U}(t)]_{\rho \sigma} |^2 .  \label{qm_prob}
\end{equation}
Typically, the oscillation probability is expressed in terms of the baseline $L$ between the source and detector; with $c=1$, then the travel time and baseline are related via $L = t$.    Using  Eq.~(\ref{qm_prob}), we explicitly compute the vacuum oscillation probability in terms of  the elements of the mixing matrix and the neutrino masses
\begin{eqnarray}
\mathcal{P}_{\sigma \rho}(L) &=& \delta_{\sigma \rho}-4 \sum^3_{\genfrac{}{}{0pt}{}{j >
k}{j,k=1}} \mathrm{Re} \left[C_{jk}^{\sigma \rho}\right] \sin^2 (\varphi_{jk}) \nonumber \\
&& +2 \sum^3_{\genfrac{}{}{0pt}{}{j > k}{j,k=1}} \mathrm{Im} \left[C_{jk}^{\sigma \rho}\right] \sin(2 \varphi_{jk})\,\,,  \label{pab_vac}
\end{eqnarray}
with $C_{jk}^{\sigma \rho} := U_{\sigma j} U^*_{\sigma k} U_{\rho k} 
U^*_{\rho
j}$ and $\varphi_{jk} := \Delta_{jk} L/(4E)$, where the neutrino mass-squared differences are
$\Delta_{jk} := m_j^2 - m_k^2$.  It is worth noting that the oscillation probability depends on only the {\em differences} in the square of the mass eigenstates.  Again, this reflects the fact that adding a multiple of the identity to the Hamiltonian in Eq.~(\ref{vac_ode}) does not impact the oscillation probability.

A general $3\times 3$ unitary matrix can be parametrized with nine real parameters:  three mixing angles, which would parametrize an orthogonal matrix, and six phases.  Not all of these phases are physically meaningful, and in fact, only one phase is of consequence in three-flavor oscillation.  Hence, to describe the oscillation of three neutrinos, the mixing matrix $U$ can be parametrized in terms of four real quantities;  one convenient parametrization is
\begin{equation}
U = U_1(\theta_{23}) D_\delta U_2(\theta_{13}) U_3(\theta_{12}) \, , \label{mixing_matrix}
\end{equation}
where $U_j(\theta)$ is a proper rotation by angle $\theta$ about the $j$-th axis and $D_\delta = \mathrm{diag}(1,1,e^{i\delta})$ \cite{peres_2004}.  This is different from, but equivalent to, the standard parametrization found in Ref.~\cite{pdg}.   It is the goal of neutrino oscillation experiments to measure the six independent parameters  which characterize neutrino oscillations: three mixing angles $\theta_{jk}$, the $CP$ phase $\delta$, and two of the mass-squared differences $\Delta_{jk}$.    Present values for the parameters can be found in   Ref.~\cite{Capozzi:2013csa,Forero:2014bxa},  global analyses of the world's data.

Matter, even if it is relatively dense, is largely transparent to neutrinos.  Despite this fact, the presence of background matter can modify the neutrino oscillation probability   \cite{ms,w}.  Neutrinos forward scatter off the background matter through either the charged-current or neutral-current weak interaction.  The forward scattering amplitude mediated by the neutral current is independent of neutrino flavor.  To account for this interaction, we add to the vacuum Hamiltonian an effective potential, but since this potential is merely a multiple of the identity, it will not impact the neutrino oscillation probability.  On the other hand, only {\em electron} (anti-)neutrinos can forward scatter off the background electrons via the charged current; these charged current interactions do impact the oscillation probability.  
 We include this effective potential in the evolution equation 
\begin{equation}
i \partial_t \nu = \left[ \frac{1}{2E} U \mathcal{M} U^\dagger + \mathcal{V}(x) \right]\nu  . \label{mswev}
\end{equation}
The operator $\mathcal{V}(x) = \mathrm{diag}(V(x), 0,0)$ exclusively acts on the electron flavor  with a magnitude $V= \sqrt{2} G_F N_e(x)$, where $G_F$ is the Fermi coupling constant  and $N_e$ is the local electron number density.   We note that for anti-neutrinos, we need to change the algebraic sign of this 
potential and the CP phase $\delta$.  

In matter of constant density, the Hamiltonian is independent of position $H = \frac{1}{2E} U \mathcal{M} U^\dagger + \mathcal{V}$, and the time evolution of the neutrino state is, again, simply $\nu(t) = \exp[-i H t] \nu(0)$.  To actually compute the time evolution operator, we recall that in the vacuum case it was useful to shift to the mass eigenstate basis, Eq.~(\ref{time_evol_vac}).
The same construction holds in matter if we construct a set of {\em effective} mass states by diagonalizing the matter Hamiltonian.  
The eigenvalues of the matter Hamiltonian $H$ are related to the {\em effective} masses in matter $\tilde{m}_j$, and the eigenvectors form the effective mixing matrix relating these states to the flavor basis.  Effective mixing angles can be extracted from this mixing matrix $\tilde{U}$ \cite{Ohlsson:1999}, though in practice this is not necessary since the oscillation probability in constant density matter can be determined from Eq.~(\ref{pab_vac}) using only $\tilde{U}_{jk}$ and $\tilde{\Delta}_{jk}$.

Numerical subroutines which effect the diagonalization of the matter Hamiltonian are sufficient tools for phenomenologists wishing to model neutrino oscillation experiments. On the other hand, if one wishes to study neutrino propagation through matter with a arbitrary variable density profile, compact analytical expressions for the time evolution operator $\mathcal{U}(t)$ are advantageous.   To arrive at a tractable analytical expression, one must  simplify the infinite sum of products of the Hamiltonian involved in the exponential.  This has been effected in Refs.~\cite{Ohlsson:1999,Ohlsson:2000} by applying the Cayley-Hamilton theorem.  Assuming the neutrino propagates through matter of constant density, the authors express the time evolution operator as the linear combination of three matrices--the identity, the Hamiltonian, and the square of the Hamiltonian.   We take a different tack and arrive at an equivalent expression for the time evolution operator expressed as a linear combination of  the identity and the Gell-Mann matrices.  

Our expression for the time evolution operator in constant density matter is based upon one of the parameterizations of an element of $SU(3)$ found in Ref.~\cite{macfarlane}.  Since we are only concerned with oscillation physics, our Hamiltonian can be made tracelss, and it is, of course, Hermitian.  As such it can  be written in terms the Gell-Mann matrices $\lambda_j$ with $j=1,\cdots,8$ which span the Lie algebra $\mathfrak{su}(3)$.  The Hamiltonian is the generator of time translations; upon exponentiation, we arrive at the time evolution operator, an element in  fundamental representation of $SU(3)$.  

Generally, we decompose the Hamiltonian into a linear combination of Gell-Mann matrices $H := c_j \lambda_j$, where summation over $j=1,\cdots,8$ is implied.  The coefficients $c_j$ are real and can be easily computed by exploiting the product rule for the Gell-Mann matrices
\begin{equation}
\lambda_j \lambda_k = \frac{2}{3} \delta_{jk} +(d_{jk\ell}+ i f_{jk\ell}) \lambda_\ell \,,  \label{lam_prod}
\end{equation}
where the totally symmetric tensor is found from the anti-commutator of Gell-Mann matrices $d_{jk\ell} = \frac{1}{4} \mathrm{tr} [\{ \lambda_j,\lambda_k \} \lambda_\ell]$ and the totally antisymmetric structure constants are determine by the commutator  $f_{jk\ell} = \frac{1}{4i} \mathrm{tr} [[ \lambda_j,\lambda_k] \lambda_\ell]$.  Tracing over the product of $H$ and a Gell-Mann matrix isolates one of the coefficients $c_j = \frac{1}{2} \mathrm{tr}[H \lambda_j]$.  

Exponentiating this operator yields the time evolution operator $\mathcal{U}(L) = \exp[-iHL]$.  
Following Ref.~\cite{macfarlane}, we aim to decompose the time evolution operator as a linear combination of the identity and Gell-Mann matrices 
\begin{equation}
\mathcal{U}= u_0 \mathbbm{1} + i u_j \lambda_j \label{u_decomp}
\end{equation}
where $u_0$ and $u_j$ are expressed in terms of the Hamiltonian and baseline.  Using the product rule in Eq.~(\ref{lam_prod}), we find 
\begin{equation}
u_0  = \frac{1}{3} \mathrm{tr}[\mathcal{U}] , \qquad u_j = \frac{1}{2i} \mathrm{tr}[\mathcal{U}\lambda_j].
\end{equation} 
The first coefficient can be simply expressed in terms of the eigenvalues of the Hamiltonian, which we denote as $\gamma_\sigma$ with $\sigma =1,\cdots,3$.  Then the eigenvalues of $\mathcal{U}$ are simply $e^{-i\gamma_\sigma L}$, and its trace is just the sum of these
\begin{equation}
u_0 = \frac{1}{3} \sum_{\sigma=1}^3 e^{-i\gamma_\sigma L}.  \label{u0}
\end{equation}
To determine the other coefficients, we note $\frac{\partial}{\partial c_j} \mathcal{U}  = -i L \,  \mathcal{U}\lambda_j$
so that 
\begin{equation}
u_j = \frac{1}{2L} \frac{\partial}{\partial c_j} \mathrm{tr}[\mathcal{U}]  =- \frac{i}{2}\sum_{\sigma=1}^3 e^{-i\gamma_\sigma L}  \frac{\partial \gamma_\sigma}{\partial c_j} .  \label{uj1}
\end{equation}
To express the derivatives of the eigenvalues in terms of invariants of the Hamiltonian, we turn to its characteristic equation
\begin{equation}
\gamma_\sigma^3  +  \frac{1}{2} (\mathrm{tr}[H]^2 -\mathrm{tr}[H^2] )\gamma_\sigma  - \det[H] = 0
\end{equation}
where we may write $\frac{1}{2} (\mathrm{tr}[H]^2 -\mathrm{tr}[H^2] ) = -c_j c_j =: - |c|^2 $ and $\det[H] = \frac{2}{3} d_{jk\ell} c_j c_k c_\ell$, summation implied \cite{macfarlane}.  
Differentiating the characteristic equation with respect to $c_j$ and solving for $\partial \gamma_\sigma/\partial c_j$ yields
\begin{equation}
\frac{\partial \gamma_\sigma}{\partial c_j} =  \frac{2(\gamma_\sigma c_j + [c*c]_j )}{3\gamma_\sigma^2 -|c|^2} \,,
\end{equation}
where we define the (eight-component) vector $[c*c]_j:= d_{jk\ell} c_k c_\ell$.
Inserting this into Eq.~(\ref{uj1}), we finally arrive at
\begin{equation}
u_j = -i \sum_{\sigma=1}^3  \frac{e^{-i\gamma_\sigma L}}{3\gamma_\sigma^2 -|c|^2}( \gamma_\sigma c_j + [c*c]_j). \label{uj}
\end{equation}
Hence we arrive at an expression where the time evolution operator in matter of constant density can be expressed as a linear combination of the identity and Gell-Mann matrices, Eq.~(\ref{u_decomp}).
This representation will be useful when considering neutrino baselines with a piecewise-constant density profile.

\section{Parametric resonance}

We return to the more general situation in which neutrinos travel through matter with a varying density, but restrict our study to situations in which this density  varies periodically.    If the periodic density profile satisfies certain conditions, then the flavor oscillation probability can be parametrically enhanced.   We will examine the possibility of full parametric resonance in a general three-neutrino framework.

We return to Eq.~(\ref{mswev}) to describe evolution through matter with a varying  density profile with periodicity $L$, i.e., $\mathcal{V}(x+L) = \mathcal{V}(x)$.  The Hamiltonian then is periodic and can be made locally traceless throughout the neutrino's trajectory.  With a Hamiltonian specified, we can solve Eq.~(\ref{mswev}) and determine the time evolution operator through one period which we denote as  $\mathcal{U}_L := \mathcal{U}(L)$.   Since the Hamiltonian is locally traceless, then $\mathcal{U}_L$ is unitary with unit determinant; as such, there exists a matrix $C = c_j \lambda_j$, with real $c_j$, such that $\mathcal{U}_L = \exp[-iCL]$.  As above, we can decompose the time evolution operator as in Eq.~(\ref{u_decomp}) with the coefficients $u_0$ and $u_j$, Eqs.~(\ref{u0}) and (\ref{uj}), written in terms of the eigenvalues, $\gamma_\sigma$, of $C$.  Evolution through $n$ periods is simply the product of these evolution operators $\mathcal{U}(n L) = [\mathcal{U}_L]^n = \exp[-in CL]$ which can be expressed as
\begin{equation}
\mathcal{U}(nL)  =\frac{1}{3} \sum_\sigma e^{-i n \gamma_\sigma L} \mathbbm{1} + \sum_{\sigma=1}^3  \frac{e^{-in \gamma_\sigma L}}{3\gamma_\sigma^2 -|c|^2}( \gamma_\sigma c_j + [c*c]_j) \lambda_j.
\label{unl}
\end{equation}

Our interest is in the conditions on the density profile that will result in the oscillation probability $\mathcal{P}_{e\mu} \to 1$ after $n$ periods for a general set of mixing angles and mass-squared differences.  Beginning with an electron neutrino $\nu(0) = \nu_e$, the oscillation probability to a muon neutrino is merely $\mathcal{P}_{e\mu} (nL) = | [\mathcal{U}(nL)]_{21}|^2$.  Exploiting the unitarity of the time evolution operator, we find that parametric resonance is achieved whenever $[\mathcal{U}(nL)]_{11} =0$ and  $[\mathcal{U}(nL)]_{31}=0$.  In what follows, we will seek conditions on the baseline that can effect these conditions.

\subsection{Case (i): $\theta_{12} = \theta$, $\theta_{13} =0$, $\theta_{23}=0$ \label{casei}}

For simplicity, let us first consider two-neutrino mixing within a three-neutrino framework.  This limited case will reproduce the results of Ref.~\cite{akhmedov_1,akh_long,floquet}.
To effect a two-neutrino scenario, we suppose that there is only one nonzero mixing angle:  $\theta_{12}=\theta$ with $\theta_{13} = \theta_{23} =0$.  In this case, the Hamiltonian in matter (before zeroing the trace) takes a block diagonal form
\begin{equation}
H = \frac{1}{2E}\left(\begin{array}{ccc}  c_\theta^2 m_1^2 + s_\theta^2 m_2^2 + 2EV & c_\theta s_\theta(m_2^2 - m_1^2) & 0\\
c_\theta s_\theta(m_2^2 - m_1^2)  & c_\theta^2 m_2^2 + s_\theta^2 m_1^2& 0\\
0 & 0 & m_3^2
\end{array}  \right)
\end{equation}
where $c_\theta = \cos \theta$ and $s_\theta= \sin \theta$. The Hamiltonian is spatially dependent by virtue of the spatial dependence of the potential $V(x)$.  Upon exponentiating this Hamiltonian locally, the block structure is maintained since the matrices $\{\mathbbm{1},\lambda_1,\lambda_2,\lambda_3,\lambda_8\}$ form a subalgebra under matrix multiplication.   The time evolution through $n$ periods can thus be decomposed as $\mathcal{U}(nL) = u_0 \mathbbm{1} +i u_j \lambda_j$ with $u_j = 0$ for $j=4,\cdots, 7$. We can thus conclude that, in writing $\mathcal{U}_L =\exp[-iCL]$, many of the coefficients vanish when decomposing $C$; namely, $c_j=0$ for $j=4,\cdots,7$.

Given this, the eigenvalues of $C$ are easy to compute: $\gamma_{1,2} = \frac{1}{\sqrt{3}} c_8 \pm \sqrt{c_1^2+c_2^2+c_3^2}$ and $\gamma_3 = -\frac{2}{\sqrt{3}}c_8$.  To determine the time evolution operator through one period, $\mathcal{U}_L$, we express the matrix invariant $|c|^2$ in terms of the eigenvalues of $C$
\begin{equation}
|c|^2 = - \frac{1}{2} (\mathrm{tr}[C]^2 -\mathrm{tr}[C^2] )  = -(\gamma_1 \gamma_2 + \gamma_1 \gamma_3 + \gamma_2 \gamma_3).
\end{equation}
This simplifies the terms in the denominator of Eq.~(\ref{uj}), since we can express such terms as the product of the difference of $C$'s eigenvalues; e.g., 
\begin{equation}
3 \gamma_1^2 - |c|^2 = ( \gamma_1-\gamma_2)(\gamma_1 - \gamma_3) .
\end{equation} 
In the two-neutrino case, with $c_j=0$ for $j=4,\cdots,7$, we find $[ c* c]_j = \frac{2}{\sqrt{3}} c_8 c_j = - \gamma_3 c_j$, for $j=1,2,3$.  This simplifies the expression for $u_j$ considerably
\begin{equation}
u_j =  \frac{c_j e^{i\frac{\gamma_3}{2}nL}}{\sqrt{c_1^2+c_2^2+c_3^2}}   \sin \left( nL\sqrt{c_1^2+c_2^2+c_3^2} \right) \label{uj_2nu}
\end{equation}
for $j=1,2,3$. 

In this two-neutrino case, the parametric resonance condition requires $[\mathcal{U}(nL)]_{11} = [\mathcal{U}(nL)]_{22} =0$. The difference of these components forces $u_3=0$. From Eq.~(\ref{uj_2nu}), we see, as a consequence, that $c_3$ must vanish.   Implementing this condition,   the $\nu_e \to \nu_\mu$ amplitude is 
\begin{equation}
[\mathcal{U}(nL)]_{21} = i e^{i \frac{\gamma_3}{2}nL}\frac{c_1+ic_2}{\sqrt{c_1^2+c_2^2}} \sin\left(nL \sqrt{c_1^2+c_2^2}\right)
\end{equation}
so that $\mathcal{P}_{e\mu} (nL) = \sin^2 \left(nL \sqrt{c_1^2+c_2^2}\right)$ which can rise to unity. 
We note that setting $c_3=0$ is simply the two-neutrino parametric resonance condition previously determined in Refs.~\cite{akhmedov_1,akh_long,floquet}.

The generic two-neutrino parametric resonance condition of Refs.~\cite{akhmedov_1,akh_long,floquet} can be implemented by the proper choice of baselines and densities for a castle-wall density profile in which
\begin{equation}
V(x) = \left\{ \begin{array}{ll}
V_a, & 0 \le x \le L_a\\
V_b, & L_a < x \le L_a +L_b 
\end{array} \right.
\end{equation}
with $V(x) = V(x+L)$ where the period is $L=L_a+L_b$.  We shall replicate these results in the three-neutrino framework.

To begin, we need to compute $u_8$
\begin{equation}
u_8 =  -i \sum_{\sigma=1}^3  \frac{e^{- i\gamma_\sigma nL}}{3\gamma_\sigma^2 -|c|^2}( \gamma_\sigma c_8 + [c*c]_8).  \label{u8_gen}
\end{equation}
Recalling that in the two-neutrino case  $c_j =0$ for $j = 4, \cdots, 7$, we find the component $[c* c]_8  = \frac{1}{\sqrt{3}}(c_1^2 + c_2^2+c_3^2-c_8^2) = \frac{1}{\sqrt{3}} (|c|^2 -\frac{3}{2}\gamma_3^2).$  Focusing upon the $\sigma = 3$ term in the sum, Eq.~(\ref{u8_gen}), inserting the previous expression yields $( \gamma_3 c_8 + [c*c]_8) =-\frac{1}{\sqrt{3}}(3 \gamma_3^2  - |c|^2)$. 
 For the $\sigma = 1,2$ factors, they simplify to  $( \gamma_\sigma c_8 + [c*c]_8) =\frac{1}{2\sqrt{3}} (3 \gamma_\sigma^2  -|c|^2  ) $.
Putting this together, we arrive an expression for $u_8$ in terms of the eigenvalues of $C$
\begin{equation}
u_8 =  - \frac{i}{2\sqrt{3}}[ e^{-i\gamma_1nL} +e^{-i \gamma_2nL} -2 e^{-i\gamma_3nL}].  \label{u8}
\end{equation}
This is the value of $u_8$ after propagation through $n$ periods of the castle-wall potential; however,  similar expressions hold, {\em mutatis mutandis}, when considering propagation through one density layer.

The time-evolution operator through one period is given by $\mathcal{U}_L = \exp[-iH_bL_b]\exp[-iH_aL_a]$ where $H_{a,b} =  H_0 + \mathcal{V}_{a,b}$.  We make the usual decomposition for the time evolution operators through one of the density regions $\exp[-iH_b L_b] = w_0 \mathbbm{1} + i w_j\lambda_j$ and $\exp[-iH_a L_a] = v_0 \mathbbm{1} + i v_j\lambda_j$. 
Parametric resonance can be achieved if the following element vanishes 
\begin{equation}
u_3 = v_3 (w_0+ i \frac{1}{\sqrt{3}}  w_8) + w_3(v_0  + i \frac{1}{\sqrt{3}} v_8).
\end{equation}  
The terms in parentheses can be simplified by making use of Eq.~(\ref{u8}); e.g., $v_0 + i \frac{1}{\sqrt{3}} v_8= \exp[i\frac{\alpha_3}{2}L_a] \cos[  \frac{\alpha_1-\alpha_2}{2}L_a] $ where $\alpha_\sigma$ represent the eigenvalues of $H_a$.   The difference in the two eigenvalues $\alpha_{1,2}$  is proportional to the effective mass-squared difference $\tilde \Delta_a$  in matter of density $V_a$.  Recalling the expression  for $v_3$ from Eq.~(\ref{uj_2nu}), we identify $a_3/\sqrt{a_1^2 +a_2^2 +a_3^2}$ with $\cos 2 \tilde \theta_a$, where $\tilde \theta_a$ is the effective matter mixing angle in the region with potential $V_a$. 
 Setting $u_3=0$ to implement the parametric resonance results in the following condition for the castle-wall potential 
\begin{equation}
\cos 2\tilde{\theta}_b\, \cos \tilde \varphi_a  \sin \tilde \varphi_b  +\cos 2\tilde{\theta}_a\, \cos\tilde \varphi_b \sin \tilde \varphi_a=0  \label{2nu_prc}
\end{equation}
with $\tilde \varphi_{a,b} =\tilde{\Delta}_{a,b} L_{a,b}/(4E)$. 
This is the condition found in Ref.~\cite{akhmedov_1,akh_long,floquet}.  Simple expressions exist for the effective matter mixing angle and mass-squared differences
\begin{eqnarray}
\sin 2 \tilde \theta &=& \frac{\sin 2 \theta}{\sqrt{c_{2\theta}^2(1 - E/E_R)^2 + s_{2\theta}^2}}, \\
\tilde\Delta  &=& \Delta \sqrt{c_{2\theta}^2(1 - E/E_R)^2 + s_{2\theta}^2} ,
\end{eqnarray}
where we define the MSW resonance energy to be $E_R =\Delta_{21} c_{2\theta}/(2V)$.  

For a given vacuum mixing angle and mass-squared difference, the parametric resonance condition in Eq.~(\ref{2nu_prc}) can be satisfied if the baselines $L_a$ and $L_b$ are an odd-integer multiple of one-half the oscillation wavelengths in matter.  This half-wavelength condition forces each individual term in Eq.~(\ref{2nu_prc}) to vanish since $\cos \tilde \varphi_{a,b}=0$.  For given matter densities which produce potentials $V_{a,b}$, this results in baselines $L_{a,b} =  (2n+1) 2\pi E/\tilde{\Delta}_{a,b}$.  The parametric condition can be satisfied by a host of other combinations of matter densities and baselines as well.  We will explore some of these alternatives numerically.  

To be concrete, we choose realistic values for the neutrino oscillation parameters:  $\theta = 0.59$, $\Delta_{21} =7.54\times10^{-5}$ eV$^2$, and  $\Delta_{31} =2.47\times 10^{-3}$ eV$^2$ \cite{Capozzi:2013csa}.  In order for our results to have some relevance to neutrinos which transit the Earth's interior, we set the matter density of the first region to $\rho_a  = 4.5$ g/cm$^3$ and the second to $\rho_b =11.5$ g/cm$^3$, values which are comparable to the densities of the Earth's mantle and core \cite{prem}.    With these values, the MSW resonant energies are $E_{R_a} = 85$ MeV and $E_{R_b} = 33$ MeV.   

\begin{figure}
\includegraphics[width=8.6cm]{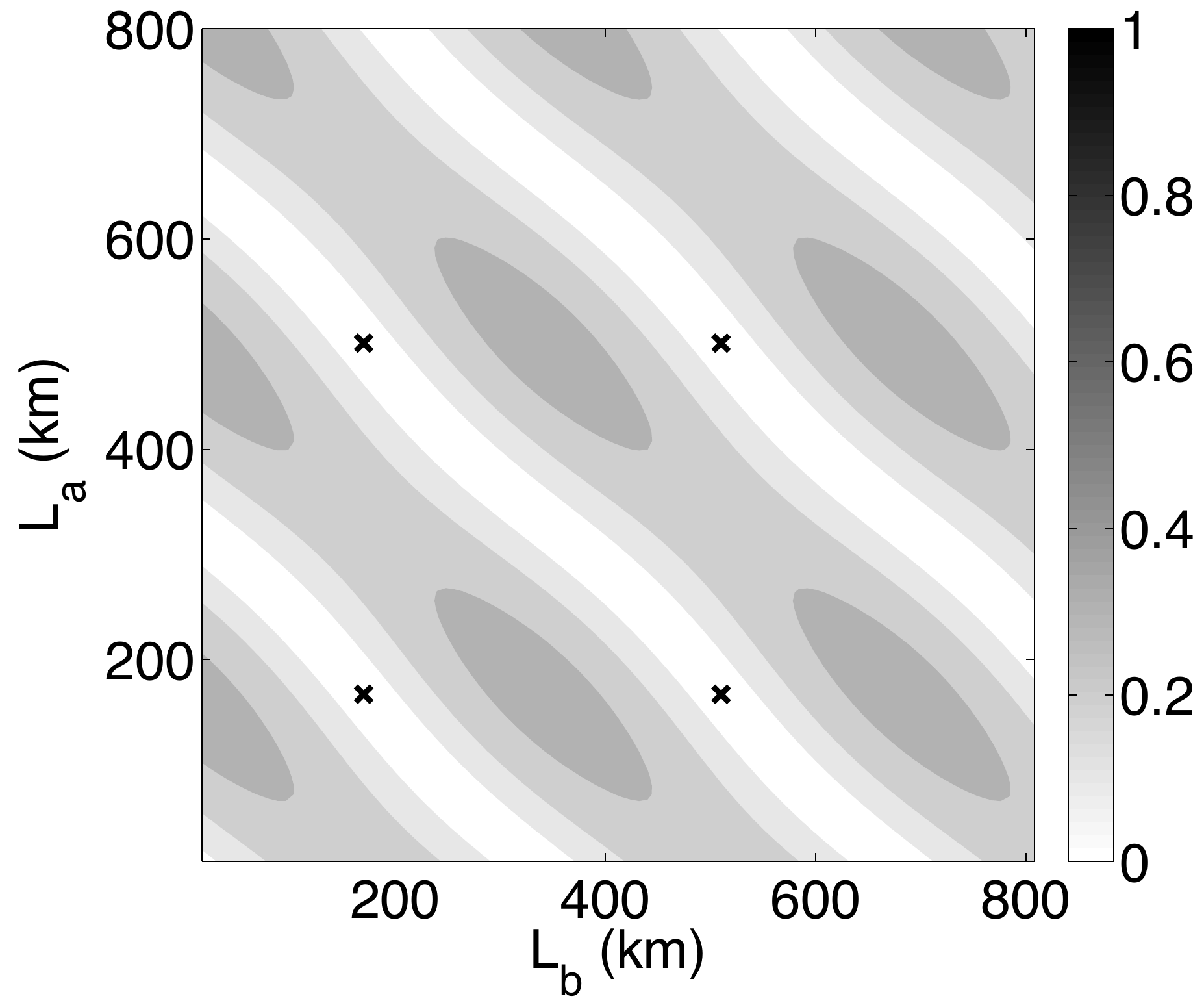}
\caption{ The modulus of $u_3$ for $E=10$ MeV and various baselines $L_{a,b}$ with densities $\rho_a =4.5$ g/cm$^3$ and $\rho_b =11.5$ g/cm$^3$  in the two-neutrino limit  $\theta=0.59$, $\Delta_{21} =7.54\times10^{-5}$ eV$^2$, and  $\Delta_{31} =2.47\times 10^{-3}$ eV$^2$. We mark the half-wavelength solutions with $\boldsymbol{\times}$. \label{fig1}}
\end{figure}

Before considering the general situation, we first examine two extreme limits analytically.  In the low energy limit, the neutrino energy is much less that the MSW resonant energy in both regions, $E \ll E_{R_{a,b}}$.  In this case, to zeroth order, the mixing angle and mass-squared difference are unchanged, so that Eq.~(\ref{2nu_prc}) becomes $\tan \tilde \varphi_a \approx - \tan\tilde  \varphi_b$ which yields a linear relationship between the acceptable baselines
\begin{equation}
L_a \approx -L_b + n \lambda_0 \,, \label{lowE}
\end{equation}
where $n$ is an integer and $\lambda_0 = 4\pi E /\Delta_{21}$, the vacuum oscillation wavelength. 
To confirm this approximation, we consider neutrinos with energy $E = 10$ MeV traveling through the castle-wall profile.  At this energy, the vacuum oscillation wavelength is 328 km, and the oscillation wavelengths in constant-density matter are $\lambda_a = 333$ km and $\lambda_b = 341$ km.  In Fig.~\ref{fig1},  we plot the modulus $|u_3|$ through one period.  The white region indicates the baselines for which $|u_3| \le 0.1$, and we mark the half-wavelength solutions with an $\boldsymbol {\times}$.  We see that the family of acceptable baselines which result in parametric resonance agree with the approximation in Eq.~(\ref{lowE}).  

\begin{figure}
\includegraphics[width=8.6cm]{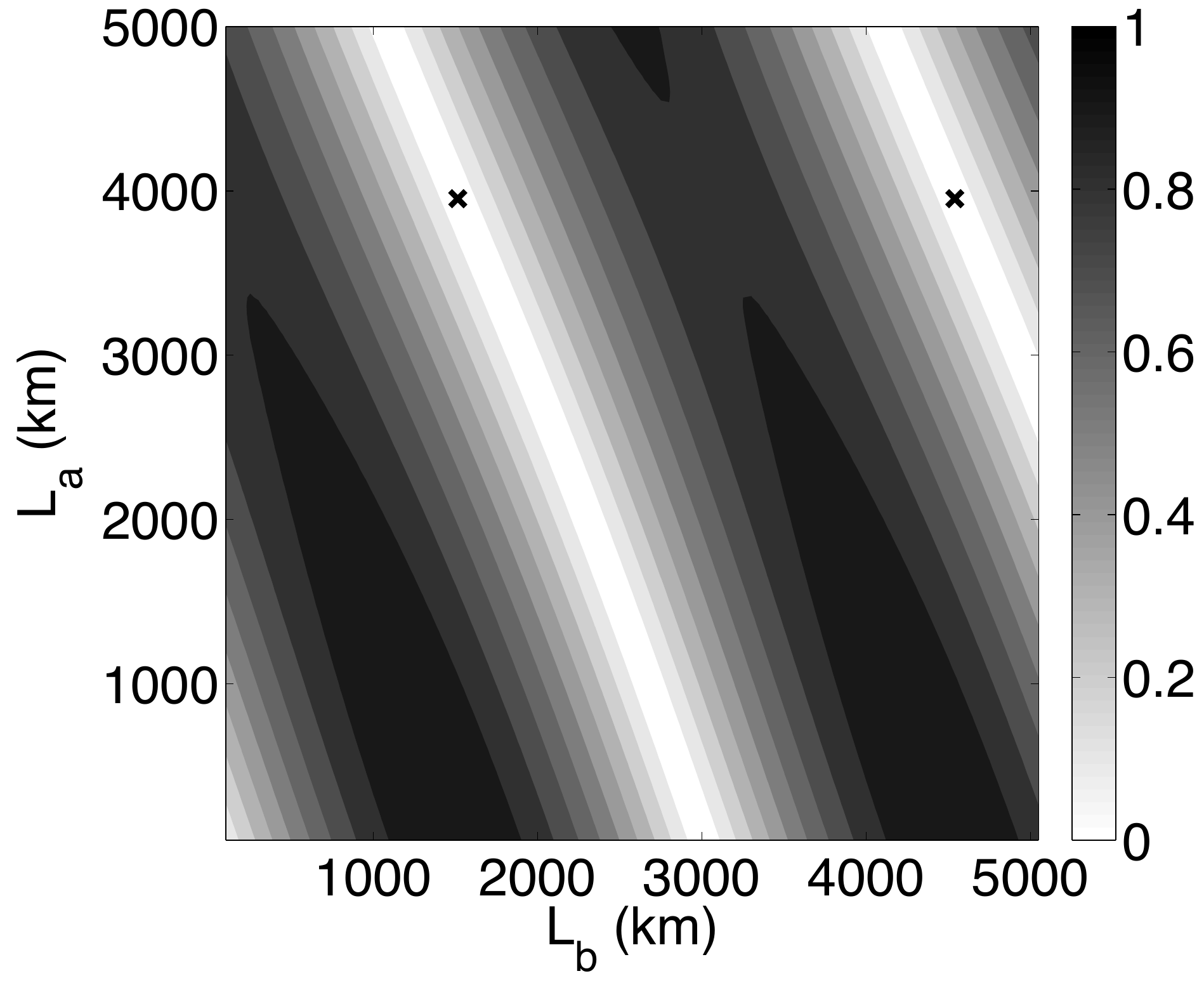}
\caption{ The modulus of $u_3$ for $E=500$ MeV and various baselines $L_{a,b}$ using the same data as Fig.~\ref{fig1}. We mark the half-wavelength solutions with $\boldsymbol{\times}$. \label{fig2}}
\end{figure}

 In the high-energy limit, the matter potential dominates the kinetic term in the Hamiltonian so that the effective mass-squared difference scales linearly with energy $\tilde \Delta \approx c_{2\theta} \Delta E/E_R$ and the effective mixing angle tends to $\pi/2$.  In this limit, the  baselines which result in parametric resonance satisfy
\begin{equation}
L_a \approx -\frac{V_b}{V_a} L_b + n \lambda_a  \,, \label{highE}
\end{equation}
where $\lambda_a \approx  2 \pi/V_a$ is the oscillation wavelength in the first region.
For the high-energy case, we consider the energy $E=500$ MeV, well beyond the MSW resonant energy for either density.  The oscillation wavelengths in matter are   $\lambda_a = 7814$ km and $\lambda_b = 2956$ km.  In Fig.~\ref{fig2}, we plot the modulus $|u_3|$ for various baselines $L_{a,b}$ at this energy.  We see that the acceptable baselines conform to the approximation in Eq.~(\ref{highE}).

\begin{figure}
\includegraphics[width=8.6cm]{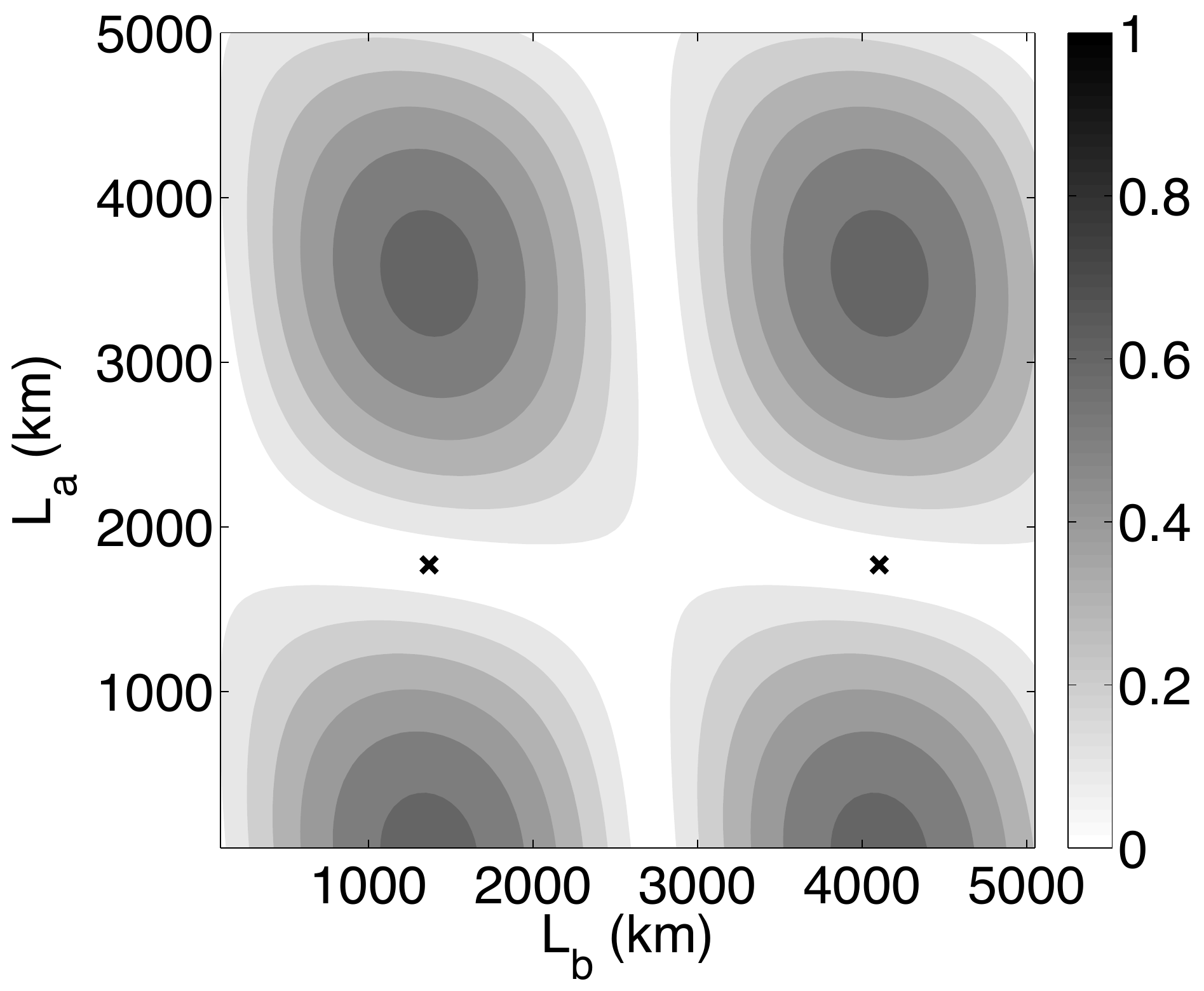}
\caption{ The modulus of $u_3$ for $E=100$ MeV and various baselines $L_{a,b}$ using the same data as Fig.~\ref{fig1}. We mark the half-wavelength solutions with $\boldsymbol{\times}$. \label{fig3}}
\end{figure}

\begin{figure}
\includegraphics[width=8.6cm]{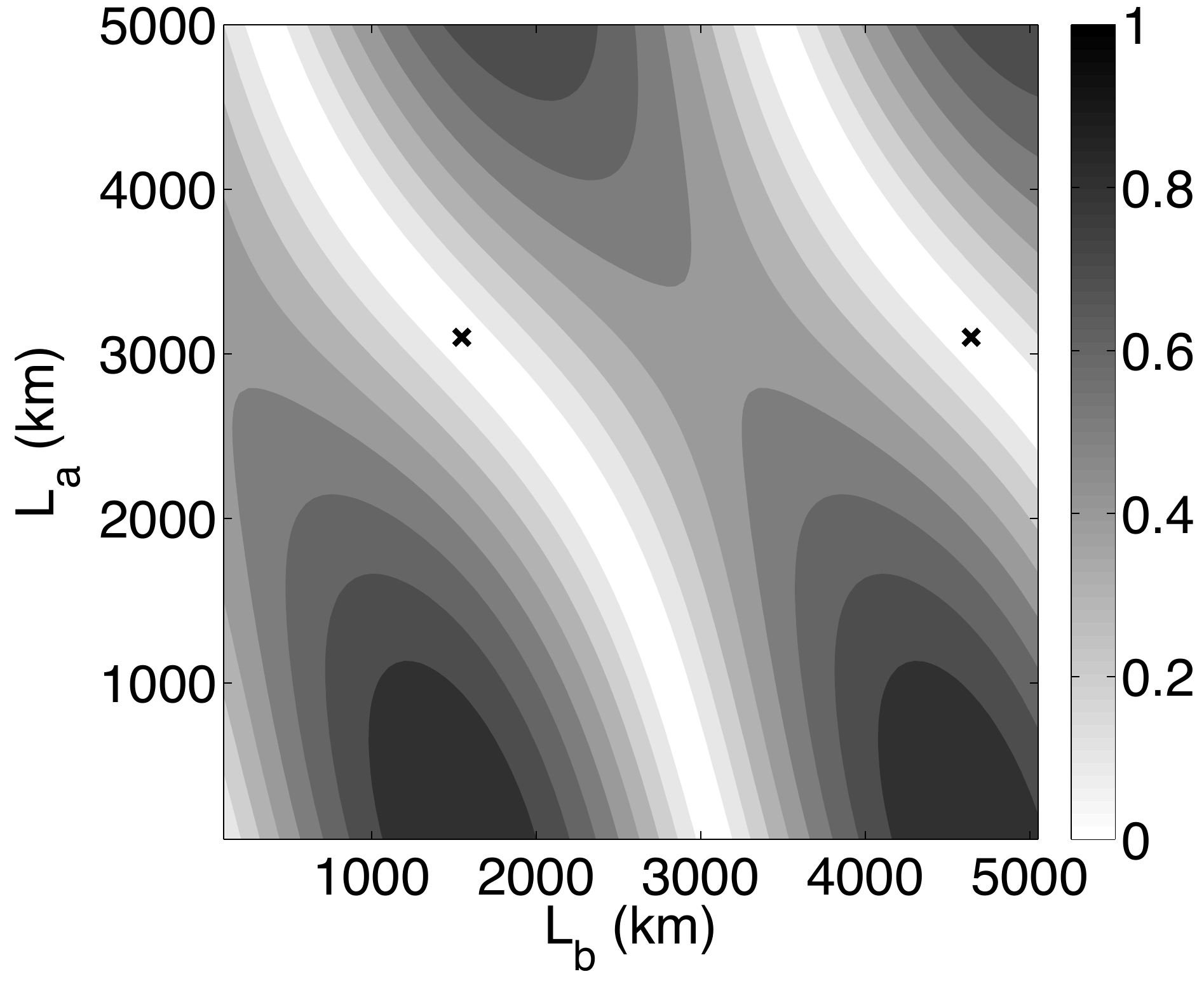}
\caption{ The modulus of $u_3$ for $E=200$ MeV and various baselines $L_{a,b}$ using the same data as Fig.~\ref{fig1}. We mark the half-wavelength solutions with $\boldsymbol{\times}$. \label{fig4}}
\end{figure}

\begin{figure}
\includegraphics[width=8.6cm]{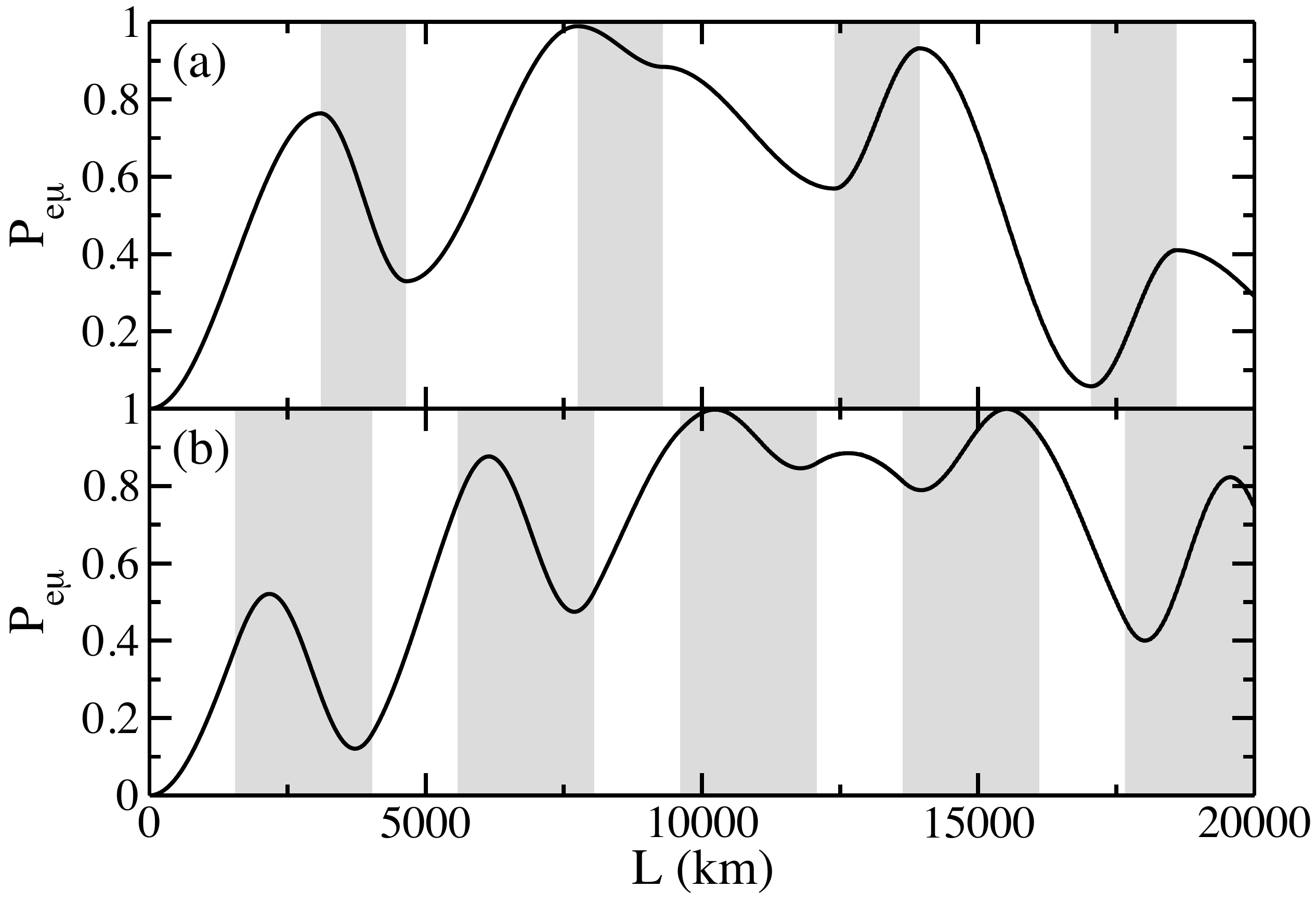}
\caption{ The oscillation probability $\mathcal{P}_{e\mu}$ for 200 MeV neutrinos through a castle-wall profile with (a) $L_a = 3100$ km and $L_b =  1548$ km and (b) $L_a = 1550$ km and $L_b = 2477$ km. The shaded regions in the plots indicate the region of density $\rho_b =11.5$ g/cm$^3$. \label{fig5}}
\end{figure}

Between the two extremes, the relationship between the allowed baselines which result in parametric resonance is much richer. We present two examples in Figs.~\ref{fig3} and \ref{fig4} for neutrino energies of 100 MeV and 200 MeV, respectively.  Focusing upon the 200 MeV case, we consider two specific castle-wall profiles which implement parametric resonance.  For this energy, the effective oscillation wavelengths in matter are $\lambda_a = 6200$ km and $\lambda_b = 3096$ km.  In Fig.~\ref{fig5}(a), we depict the oscillation probability $\mathcal{P}_{e\mu}$ for a neutrino traveling through a castle-wall density profile satisfying the half-wavelength condition.  The maximum oscillation probability in matter with a constant density $\rho_a$ is 0.76 and $\rho_b$ is 0.19, yet through parametric resonance the oscillation probability goes to unity after two periods.  In Fig.~\ref{fig5}(b), we set the length of the first region to be one-quarter oscillation wavelength $L_a = 1550$ km.   From Fig.~\ref{fig4}, we determine that if $L_b = 2477$ km then parametric resonance can be achieved.  Again, after several periods, the oscillation probability rises to unity.

\subsection{Case (ii): $\theta_{12} = \theta$, $\theta_{13} =\phi$, $\theta_{23}=0$}

We now allow for two mixing angles to be nonzero: $\theta_{12} = \theta$, $\theta_{13} =\phi$, and $\theta_{23}=0$.  
Unlike the two-neutrino case, the Hamiltonian is no longer block diagonal, and as a result, all elements $[\mathcal{U}]_{\rho \sigma}$ will generally be non-zero, permitting oscillation amongst all flavors.  
This complicates the analytical treatment of parametric resonance significantly, yet the fundamental requirement is still the same.  If the oscillation probability $\mathcal{P}_{e\mu}$ is to tend to unity for neutrinos traveling through a periodic matter density profile, then $[\mathcal{U}_L]_{11} =[\mathcal{U}_L]_{31} =0$.  If this condition holds, then unitarity also implies $[\mathcal{U}_L]_{22} =[\mathcal{U}_L]_{23} =0$.  

As with the two-neutrino case, the difference of the two diagonal elements $[\mathcal{U}_L]_{11}$ and $[\mathcal{U}_L]_{22}$ requires $u_3 = 0$.  
There are now two paths by which one can pursue parametric resonance; we will only consider one  which entails the additional requirement of $[\mathcal{U}_L]_{32} =0$. 
This implies $u_j = 0$ for $j=4,5,6,7$.   Implementing these conditions for a general periodic matter profile is intractable, so we will consider the same castle-wall profile as in the previous section with densities $\rho_a = 4.5$ g/cm$^3$ and $\rho_b = 11.5$ g/cm$^3$.  We use the same neutrino oscillation parameters as above along with $\phi = 0.15$ \cite{Capozzi:2013csa}. 

Before considering numerical results, we aim to gain a qualitative understanding of parametric resonance in this case by treating $\phi$ as a perturbative parameter.  The Hamiltonian, in constant density matter, governs the behavior of the system, so we first examine how it changes to first order in $\phi$.  To be concrete, we consider the region of density $\rho_a$ and recall our decomposition in terms of the Gell-Mann matrices $H_a = a_j \lambda_j$.  The two-neutrino portion, $a_1, a_2, a_3,$ and $a_8$, of the Hamiltonian is unchanged up to $\mathcal{O}(\phi^2)$, but there are two additional off-diagonal contributions to the Hamiltonian
\begin{equation}
a_4 \approx \tfrac{1}{2E} (c_\theta^2 \Delta_{31} + s_\theta^2 \Delta_{32})\phi, \quad a_6 \approx - \tfrac{1}{2E} c_\theta s_\theta \Delta_{21} \phi . \label{a4a6}
\end{equation}
As a result, the eigenvalues of $H$ do not change to first order in $\phi$, but since $\theta_{13}$ is non-zero, there are two independent oscillation scales given by the effective mass-squared differences in matter
\begin{eqnarray}
\tilde\Delta_{21}^a &:=& 2E(\alpha_2 - \alpha_1) \approx \Delta_{21} \sqrt{c_{2\theta}^2\left(1 - \tfrac{E}{E_{R_a}}\right)^2 + s_{2\theta}^2} \nonumber \\
\\
\tilde \Delta_{32}^a &:=& 2E(\alpha_3 -\alpha_2) \approx \tfrac{1}{2} (\Delta_{31} + \Delta_{32}) -EV -\tfrac{1}{2}\tilde \Delta_{21}^a \nonumber \\
\end{eqnarray}
with $\tilde \Delta_{31}^a = \tilde \Delta_{32}^a + \tilde \Delta_{21}^a$.  Since we are neglecting terms that are $\mathcal{O}(\phi^2)$, we should note the vast difference between oscillation scales $\Delta_{21}/ \Delta_{32} \sim 0.03$.  
Since this is much smaller than $\phi$, we will also neglect terms that are $\mathcal{O}\left(\phi \frac{\Delta_{21}}{\Delta_{32}}\right)$.  
Given this approximation, we take $a_6\approx 0$ so that the only leading order deviation from the two-neutrino Hamiltonian is due to $a_4$ which can be further approximated as $a_4 \approx \frac{1}{2E} \Delta_{32} \phi$, again neglecting a term that is  $\mathcal{O}\left(\phi \frac{\Delta_{21}}{\Delta_{32}}\right)$.

The time evolution operator through one period is $\mathcal{U}_L = \exp[-iH_b L_b]\exp[-iH_a L_a]$, and we make the usual decomposition $\exp[-iH_b L_b] = w_0 \mathbbm{1} + i w_j\lambda_j$ and $\exp[-iH_a L_a] = v_0 \mathbbm{1} + i v_j\lambda_j$.  To implement parametric resonance through this castle-wall profile, we require $u_j =0$ for $j=3,\cdots, 7$.  Given the above approximations, only $u_3$ and $u_4$ are appreciable to leading order.   Requiring $u_3=0$ results in the old two-neutrino condition for parametric resonance, Eq.~(\ref{2nu_prc}).  The new requirement $u_4 =0$ demands
\begin{equation}
w_4 \left[ v_0 -\tfrac{i}{2 \sqrt{3}} v_8 + \tfrac{i}{2} v_3 \right] + v_4  \left[ w_0 -\tfrac{i}{2 \sqrt{3}} w_8 + \tfrac{i}{2} w_3 \right]   \approx 0,  \label{u4_i}
\end{equation}
consistently applying the approximation $v_6 \approx w_6 \approx 0$.  
We examine each factor in this equation for energies greater than the resonance energy $E\gg E_{R_{a,b}}$.

We begin with the factor in square brackets in Eq.~(\ref{u4_i}).  Each term in this factor is known from the previous work with the two-neutrino case, which is valid to $\mathcal{O}(\phi)$.   Repurposing that work, we find 
\begin{eqnarray}
v_0 -\tfrac{i}{2 \sqrt{3}} v_8 + \tfrac{i}{2} v_3 &\approx& \tfrac{1}{4}[e^{-i\alpha_1 L_a}+e^{-i\alpha_2 L_a} +2 e^{-i \alpha_3 L_a}  ]  \nonumber \\
&& + \frac{a_3 }{2( \alpha_1 - \alpha_2)}[e^{-i\alpha_1 L_a}-e^{-i\alpha_2 L_a} ].\nonumber\\
\end{eqnarray}
For neutrinos with an energy on the order of hundreds of MeV, we make a further approximation, $E \gg E_{R_{a,b}}$.  In this limit, we find
\begin{eqnarray}
a_3 &\approx& \tfrac{1}{2}V_a \\
\tilde \Delta_{21}^a &\approx& 2EV_a \\
\tilde \Delta_{32}^a &\approx& \tfrac{1}{2}(\Delta_{31} + \Delta_{32}) - 2EV_a.
\end{eqnarray}
Making the appropriate substitutions, we arrive at the expression
\begin{equation}
v_0 -\tfrac{i}{2 \sqrt{3}} v_8 + \tfrac{i}{2} v_3 \approx e^{-i\frac{( \alpha_3 + \alpha_2)}{2}L_a } \cos \left(\frac{\tilde \Delta_{32}^a L_a}{4E} \right)
\end{equation}

The other factor in  Eq.~(\ref{u4_i}) is specific to the three-neutrino case.  Noting that, in this case, $[a*a]_4 = a_4\left( a_3 - \frac{1}{\sqrt{3}} a_8\right)$, we find in a region of constant density
\begin{equation}
v_4 = -i a_4 \sum \frac{e^{-i \alpha_\sigma}}{3\alpha_\sigma^2 -|a|^2} \left(\alpha_\sigma  +a_3 - \tfrac{1}{\sqrt{3}} a_8  \right).
\end{equation}
In the high energy limit, $E\gg E_{R_a}$, this becomes
\begin{equation}
v_4 \approx -\frac{4E a_4 }{\tilde{\Delta}_{32}^a} e^{-i\frac{( \alpha_3 + \alpha_2)}{2}L_a }  \sin \left(\frac{\tilde \Delta_{32}^a L_a}{4E} \right)
\end{equation}
where $a_4$ is given in Eq.~(\ref{a4a6}).  

With these two factors determined in matter of constant density, the new additional requirement for parametric resonance, Eq.~(\ref{u4_i}), in the limit  $E\gg E_{R_{a,b}}$ is approximately
\begin{equation}
\frac{1}{\tilde{\Delta}^b_{32}}   \sin \tilde \vartheta_b  \cos\tilde \vartheta_a
    + \frac{1 }{\tilde{\Delta}_{32}^a}  \sin \tilde \vartheta_a   \cos \tilde \vartheta_b  \approx 0, \label{u4_ii}
  \end{equation}
where $\tilde \vartheta_{a,b} := \tilde \Delta_{32}^{a,b} L_{a,b}/4E$.  This equation has a structure similar to the two-neutrino parametric resonance condition, Eq.~(\ref{2nu_prc}).  

In what follows, let us further restrict the neutrino energies under consideration by providing an upper bound $2E(V_b-V_a) <  \Delta_{32}$.  Given the densities under consideration, the upper bound on energies is around 4.5 GeV.  If the energy is significantly less than this upper bound, then the ratio of effective mass-squared differences $\tilde \Delta^a_{32}/\tilde \Delta^b_{32}$ is unity with corrections $\mathcal{O}\left( \frac{E(V_b-V_a)}{\Delta_{32}}\right)$.
In this limit, the additional condition for parametric resonance can be simply expressed in terms of the baselines 
\begin{equation}
L_a \approx -\frac{\tilde \Delta^b_{32}}{\tilde \Delta^a_{32}} L_b + n \frac{4\pi E}{\tilde \Delta^a_{32}} . \label{u4_iii}
\end{equation}

In summary, parametric resonance can be achieved in this three-neutrino scenario of ``small" $\phi$ if the two-neutrino parametric resonance condition, Eq.~(\ref{2nu_prc}), is satisfied along with the new constraint, Eq.~(\ref{u4_iii}).  This latter equation is valid for neutrino energies between a few hundred MeV and a few GeV.  The new constraint yields another  linear relationship between $L_a$ and $L_b$ with an absolute slope near unity.  For energies around a few hundred MeV, the family of curves generated by  Eq.~(\ref{u4_iii}) will be relatively dense since $\sin \tilde \vartheta$ oscillates rapidly at these low energies. For energies above 500 MeV, the additional constraint on parametric resonance will become appreciable.  In considering the acceptable baselines, what were once continuous regions of solutions in $L_a$-$L_b$ parameter space now become a series of isolated solutions.

\begin{figure}
\includegraphics[width=8.6cm]{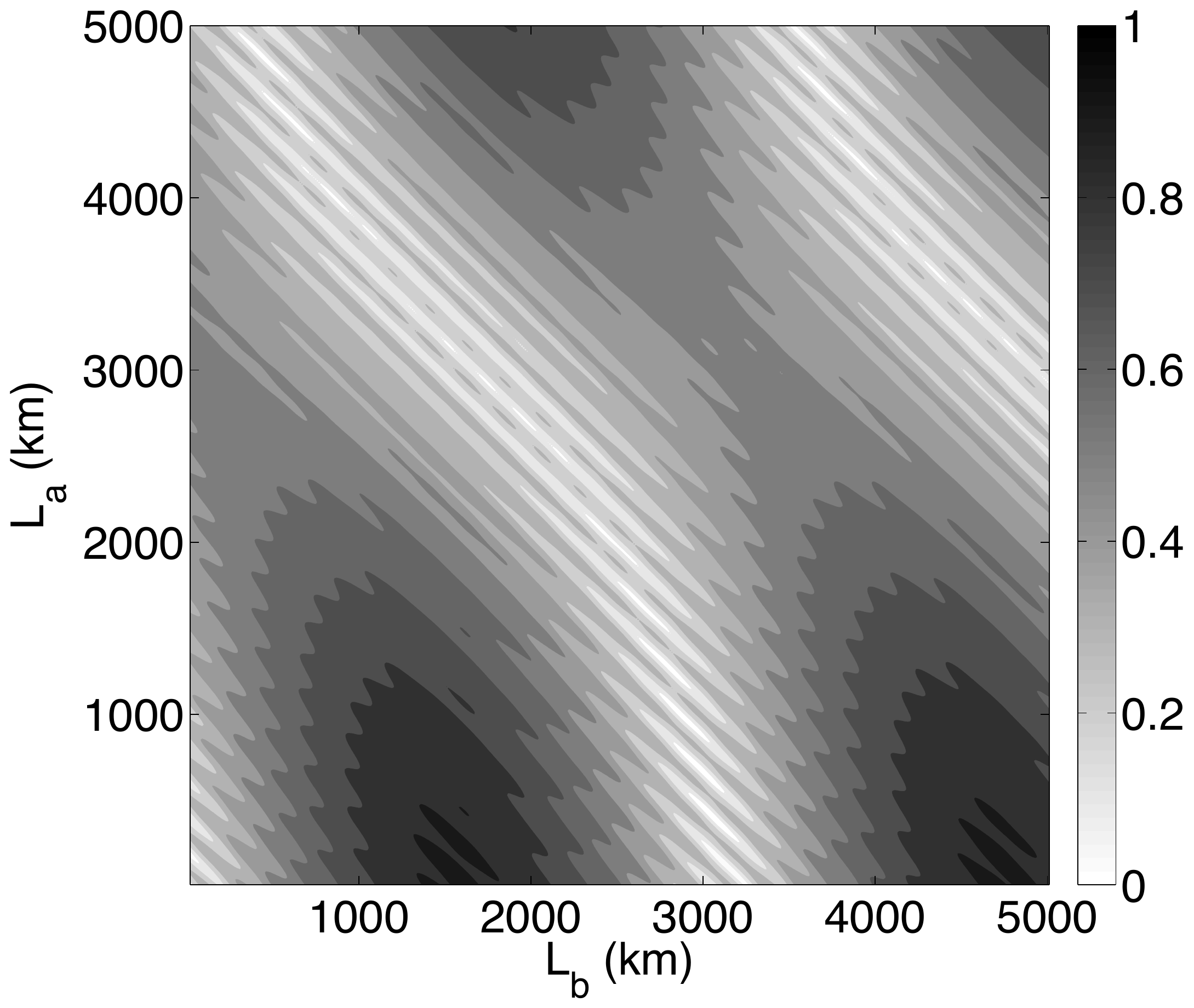}
\caption{ A plot of $\overline u$ for $E=200$ MeV and various baselines $L_{a,b}$ with densities $\rho_a =4.5$ g/cm$^3$ and $\rho_b =11.5$ g/cm$^3$  with $\theta=0.59$, $\phi=0.15$, $\Delta_{21} =7.54\times10^{-5}$ eV$^2$, and  $\Delta_{31} =2.47\times 10^{-3}$ eV$^2$.
 \label{fig6}}
\end{figure}

We will examine these results numerically.  Since parametric resonance requires $u_j = 0$ for $j=3,\cdots, 7$, 
we will aim to determine the acceptable baselines which minimize the parameter 
\begin{equation}
\overline u := \sqrt{|u_3|^2 + |u_4|^2+|u_5|^2+|u_6|^2+|u_7|^2}.
\end{equation}
In Fig.~\ref{fig6}, we find the baselines which minimize $\overline u$, resulting in parametric resonance, in the castle-wall profile for neutrinos with an energy of 200 MeV. The white regions in the plot indicate where $\overline u \le 0.1$.  Comparing this plot with its two-neutrino analog  in Fig.~\ref{fig4}, we see that the contours are predominantly determined by $u_3$, but there are interfering higher frequency contributions attributable  to oscillations dependent upon the $\Delta_{32}$ mass-squared difference.  As deduced above, the dominant interference term is $u_4$. Numerically, we determine the effective mass-squared differences $\tilde \Delta_{32}^a = 2.36\times 10^{-3}$ eV$^2$ and  $\tilde \Delta_{32}^b =2.27 \times 10^{-3} $ eV$^2$.  For this energy, we determine that the constraint derived from $u_4=0$, Eq.~(\ref{u4_iii}), becomes $L_a = -0.96\, L_b + (210\text{ km}) \, n$.  Close inspection of the plot of $|u_4|$ for various baselines (not shown) is consistent with this family of lines.  At this energy, the impact of the $u_4$ in determining the acceptable baselines is minimal, and the two-neutrino parametric resonance condition, Eq.~(\ref{2nu_prc}), represents a good approximation.

\begin{figure}
\includegraphics[width=8.6cm]{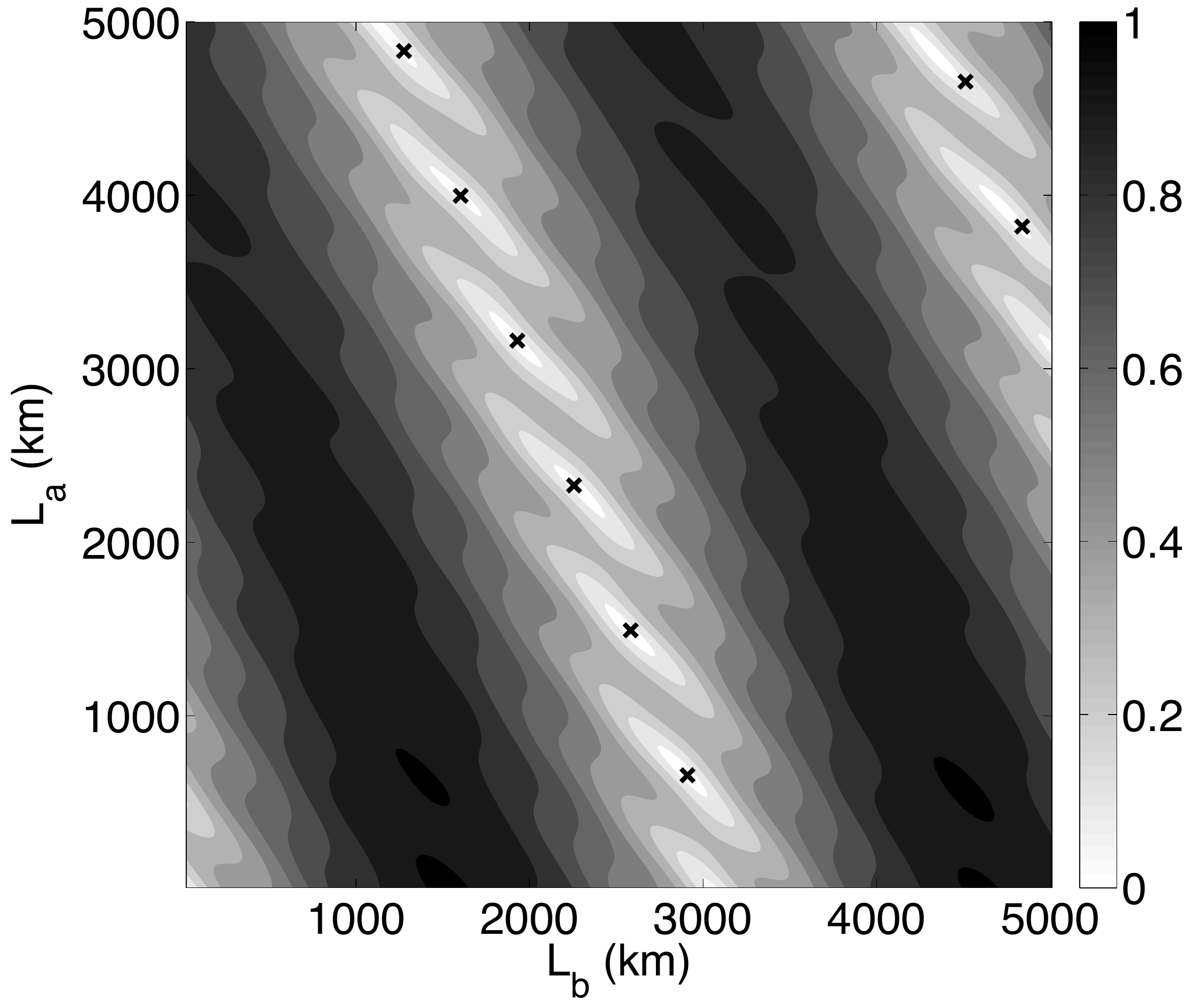}
\caption{ A plot of $\overline u$ for $E=500$ MeV and various baselines $L_{a,b}$ using the same data as Fig.~\ref{fig6}.  We mark the approximate minima of $\overline u$, solutions to the equations Eqs.~(\ref{500mevu3}) and (\ref{500mevu4}), with the symbol $\boldsymbol \times$.
 \label{fig7}}
\end{figure}

In Fig.~\ref{fig7}, we plot $\overline u$ for neutrinos with energies of 500 MeV.  In comparing the allowed baselines which result in parametric resonance with those in the two-neutrino case, Fig.~\ref{fig2}, the impact of interference between the two oscillation scales is significant.  In the two-neutrino case, setting $u_3=0$ results in the (approximate) family of allowed baselines
\begin{equation}
L_a \approx - 2.56\, L_b + (8094\text{ km}) \, n_1\, \label{500mevu3}
\end{equation}  
where we have used the numerically determined value for the wavelength relevant for oscillations due to $\tilde \Delta_{21}^a$.
But, in the three-neutrino case, the continuous region of allowed baselines becomes a series of isolated points in $L_a$-$L_b$ parameter space. With the effective mass-squared differences $\tilde \Delta_{32}^a = 2.28\times 10^{-3}$ eV$^2$ and  $\tilde \Delta_{32}^b =2.04 \times 10^{-3} $ eV$^2$, the new additional constraint, Eq.~(\ref{u4_iii}), becomes  
\begin{equation}
L_a \approx  -0.89\, L_b + (543\text{ km}) \, n_2.  \label{500mevu4}
\end{equation}
 The intersection of these two curves, denoted by $\boldsymbol \times$  in Fig.~\ref{fig7},  approximates the acceptable baselines indicated by the local minima of $\overline u$. The position of the approximate minima is not exact due to higher order terms not considered in the analytic work, but it does provide a reasonable estimate.

\begin{figure}
\includegraphics[width=8.6cm]{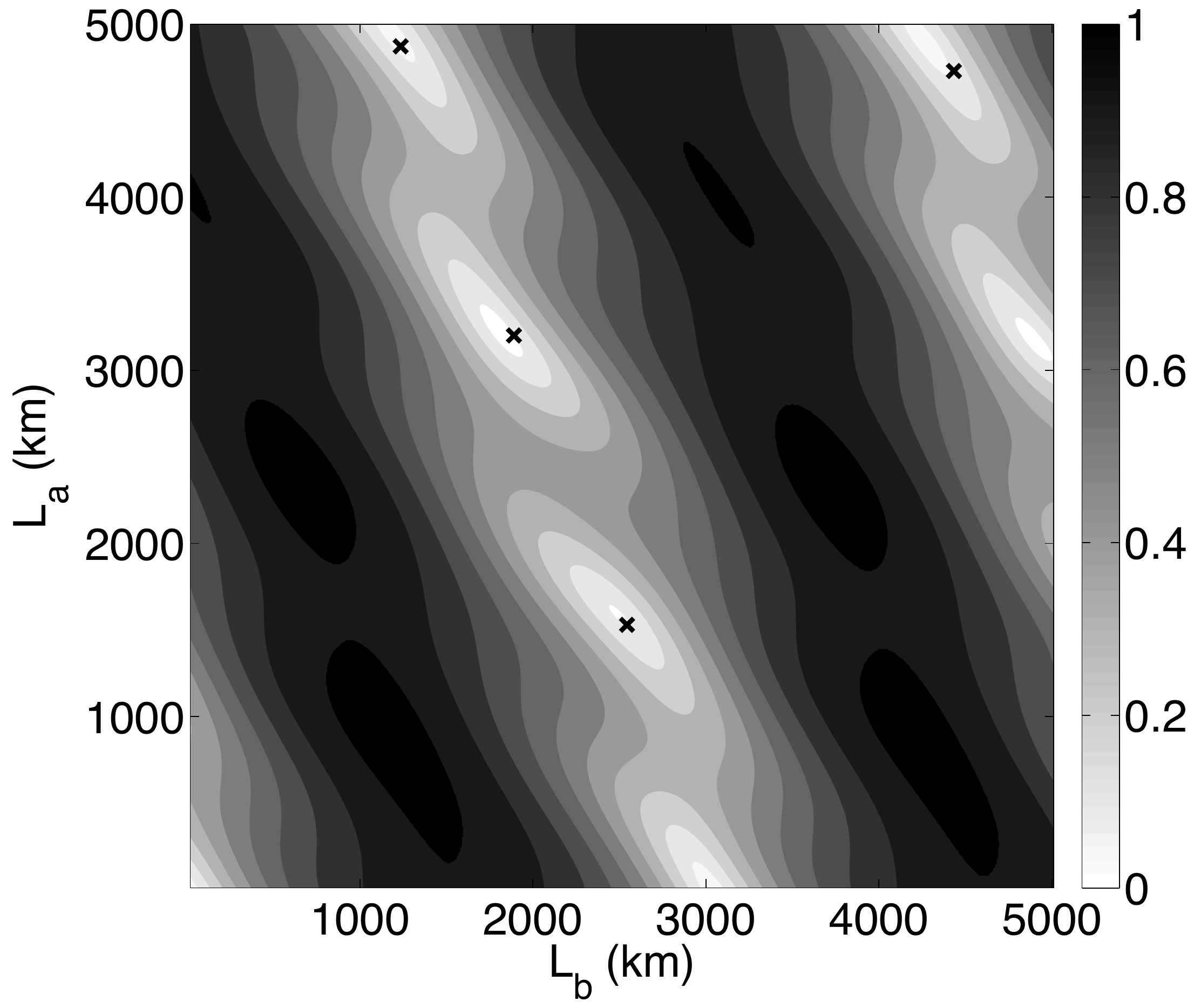}
\caption{ A plot of $\overline u$ for $E=1$ GeV and various baselines $L_{a,b}$ using the same data as Fig.~\ref{fig6}. We mark the approximate minima of $\overline u$, solutions to the equations Eqs.~(\ref{1gevu3}) and (\ref{1gevu4}), with the symbol $\boldsymbol \times$.
 \label{fig8}}
\end{figure}

We consider one more example for neutrinos with an energy of 1 GeV, Fig.~\ref{fig8}.  For such a high energy, the condition $u_3 =0$ should yield a robust linear relationship between $L_a$ and $L_b$, Eq.~(\ref{1gevu3}), while setting $u_4 = 0$ yields an approximate relationship between baselines, Eq.~(\ref{1gevu4}),
\begin{eqnarray}
L_a &\approx& - 2.56\, L_b + (8030 \text{ km}) \, n_1 \label{1gevu3} \\
L_a &\approx& -0.77 \, L_b + (1165 \text{ km}) \, n_2 \label{1gevu4} ,
\end{eqnarray}  
where the effective mass-squared differences are $\tilde \Delta_{32}^a =2.12 \times 10^{-3}$ eV$^2$ and  $\tilde \Delta_{32}^b = 1.64 \times 10^{-3} $ eV$^2$.  
 Again, the intersection of these two lines indicates the approximate position in the $L_a$-$L_b$ parameter space at which parametric resonance can be achieved; we indicate the points with the symbol $\boldsymbol \times$ in Fig.~\ref{fig8}.

\begin{figure}
\includegraphics[width=8.6cm]{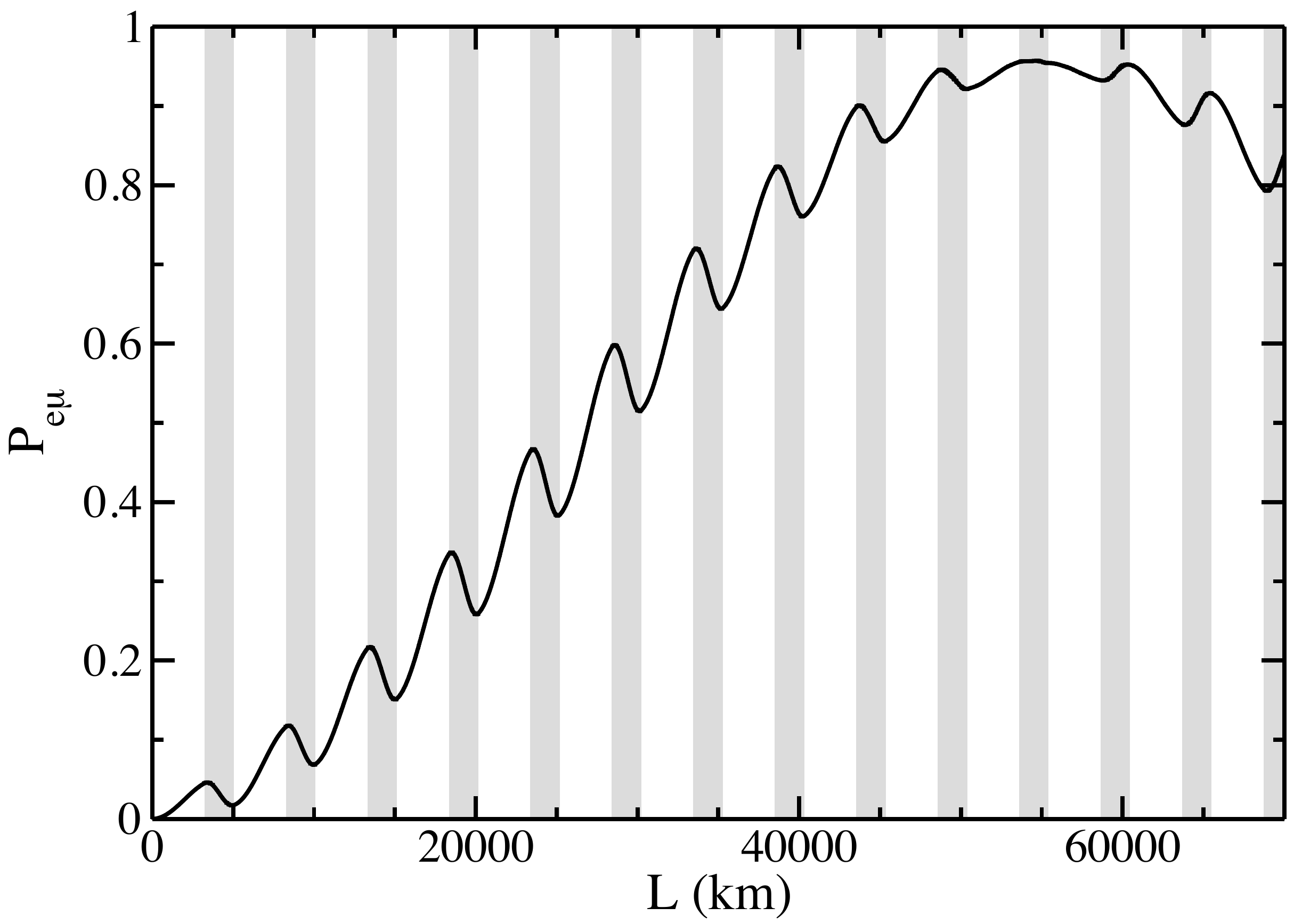}
\caption{ The oscillation probability $\mathcal{P}_{e\mu}$ for 1 GeV neutrinos through a castle-wall profile with $L_a = 3220$ km and $L_b =  1819$ km. The shaded areas indicate regions of density $\rho_b =11.5$ g/cm$^3$. \label{fig9}}
\end{figure}

For the 1 GeV case, the absolute minimum for the parameter $\overline u$ in the range of baselines shown in Fig.~\ref{fig8} is 0.06, occurring at   
$L_a = 3210$ km and $L_b = 1830$ km.   We plot the oscillation probability in Fig.~\ref{fig9} for this castle-wall profile.  Indeed, the oscillation probability does exhibit parametric enhancement; however, since the minimum of $\overline u$ does not vanish, the probability cannot go to unity but rather attains a maximum of 0.96.  Still, this is a dramatic increase over the maximum constant-density oscillation probability.  The neutrino energy of 1 GeV is well beyond the resonant energies of 85 MeV and 33 MeV, dramatically suppressing $\mathcal{P}_{e\mu}$ to a maximum value of 0.05 for travel through constant density $\rho_a$ and 0.007 for travel through $\rho_b$.   Since these constant density oscillation probabilities are so small, it takes roughly 10 periods to achieve the maximal parametric enhancement.

We remark that there are many baselines through which one can achieve parametric resonance.  Ultimately, what is required for parametric resonance is that select elements of the time evolution operator must vanish, $u_j =0$ for $j=3,\cdots, 7$.  We have implemented these conditions for a simple castle-wall profile with solutions given by $\overline u =0$.  The minima of $\overline u$ in Figs.~\ref{fig6} through \ref{fig8} show the proper baselines $L_a$ and $L_b$ that result in parametric resonance; however, other points in the $L_a$-$L_b$ parameter space with $\overline u$ significantly greater than zero are not necessarily excluded from parametric resonance.  As a case in point, consider the profile with baselines given by $L_a = 2400$ km and $L_b = 2150$ km in Fig.~\ref{fig8}.  Here, we find $\overline u = 0.44$, yet, this choice of parameters does permit significant parametric enhancement of the oscillation probability.  The reason for this is that $\overline u$ is large after one period $L$, but it does attain a much smaller value ($\overline u = 0.06$) after two periods $2L$, satisfying the condition for parametric resonance over a different period.  Our parametric resonance conditions can only highlight the appropriate baselines that will show resonance; they do not necessarily exclude other baselines.

\subsection{Case (iii): $\theta_{12} = \theta$, $\theta_{13} =\phi$, $\theta_{23}=\psi$}

We now allow $\theta_{23} = \psi$ to be nonzero.  
Recalling the parametrization for the mixing matrix $U(\theta,\phi,\psi)$, Eq.~(\ref{mixing_matrix}), we see that we can peel off the $\theta_{23}$-rotation, $U_1(\psi)$, and since it commutes with the matter potential $\mathcal{V}(x)$, we can write the neutrino evolution equation, Eq.~(\ref{mswev}), in terms of the state $\hat \nu = U_1(\psi)^\dagger \nu$
\begin{equation}
i \partial_t \hat \nu = \left[ \frac{1}{2E} U(\theta,\phi,0) \mathcal{M} U(\theta,\phi,0)^\dagger + \mathcal{V}(x) \right]\hat \nu  . 
\end{equation}
The evolution of the state $\hat \nu$ through a periodic matter profile was addressed in the previous two subsections.

\begin{figure}
\includegraphics[width=8.6cm]{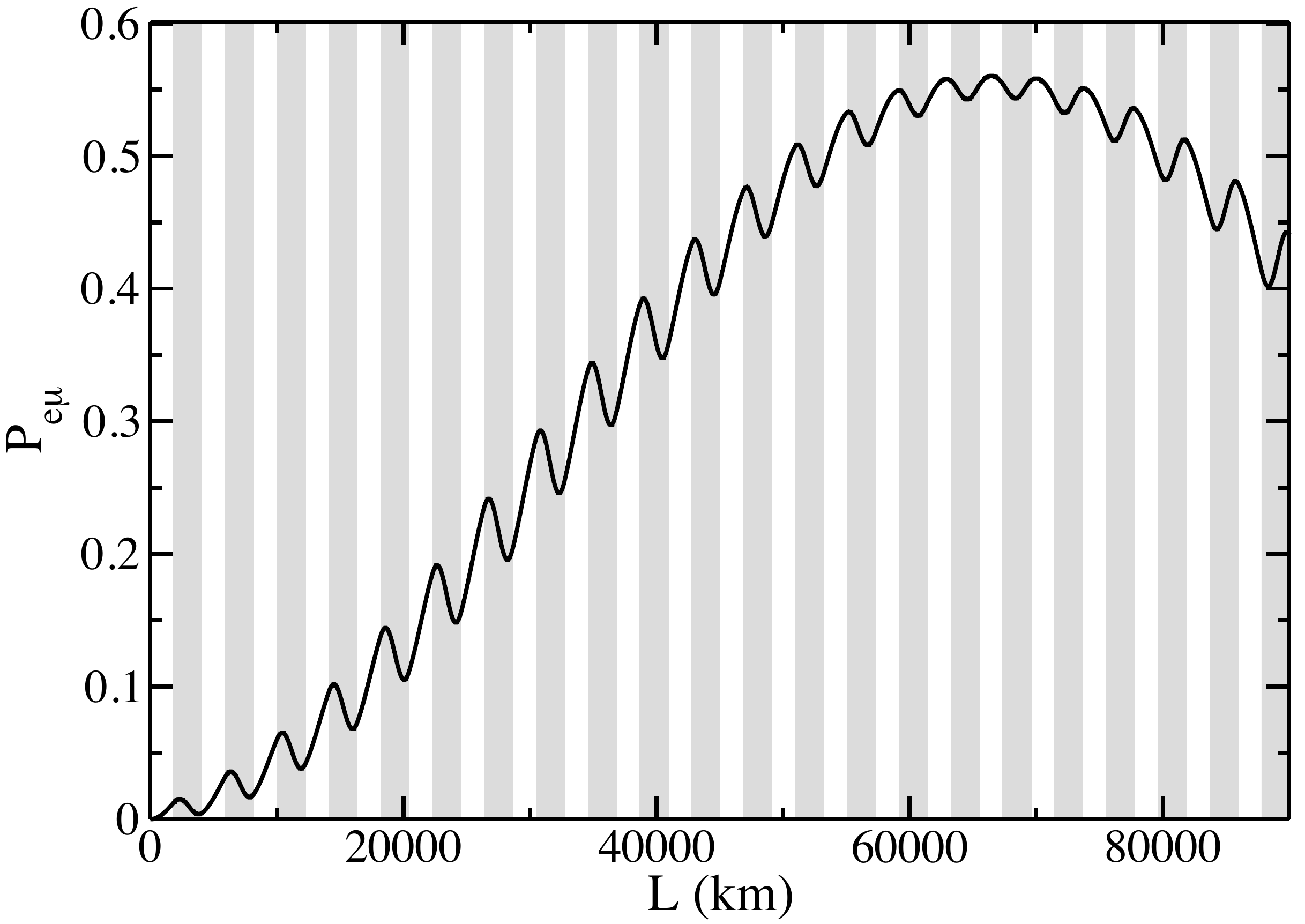}
\caption{ The oscillation probability $\mathcal{P}_{e\mu}$ for 1 GeV neutrinos through a castle-wall profile with $L_a = 1800$ km and $L_b =  2296$ km. The shaded areas indicate regions of density $\rho_b =11.5$ g/cm$^3$.  We set $\theta = 0.59$, $\phi = 0$, and $\psi = 0.72$. \label{fig10}}
\end{figure}

Specializing first to the case in which $\phi=0$, we find that the mixing angle $\psi$ places an upper bound on the maximum oscillation probability $\mathcal{P}_{e\mu}$ that can be achieved through parametric resonance.  With $\phi=0$, the evolution of the state $\hat \nu$ involves mixing amongst only two flavors as in Case (i).  Regardless of the values of the vacuum mixing angle $\theta$ or mass-squared differences, there exists a periodic matter profile that allows the oscillation probability, $\hat{\mathcal{P}}_{e\mu}$,  for the state $\hat \nu$  to reach unity; that is, for some number of periods, $| \hat{\mathcal{U}}_{21}(nL)| \to 1$.  For the true neutrino state $\nu$, we can rotate bases to relate this element of the time evolution operator to the two-neutrino one ${\mathcal{U}}_{21} =  \cos \psi \, \hat{\mathcal{U}}_{21}$.  Given this, the maximum oscillation probability achievable through parametric resonance (with $\phi=0$) is $\mathcal{P}_{e\mu} \to \cos^2 \psi$; i.e., full parametric enhancement (to unity) is not possible.  
As an example, we consider 1 GeV neutrinos with $\psi=0.72$ \cite{Capozzi:2013csa} and the usual  mixing angle $\theta=0.59$.  The condition for (partial) parametric resonance is set by the two-neutrino case.  We choose one such solution with baselines $L_a = 1800$ km and $L_b=2296$ km which yields $\hat u_3 =0$.  In Fig.~\ref{fig10}, we plot the oscillation probability through the castle-wall profile which is clearly bounded by $\cos^2 \psi = 0.57$.

\begin{figure}
\includegraphics[width=8.6cm]{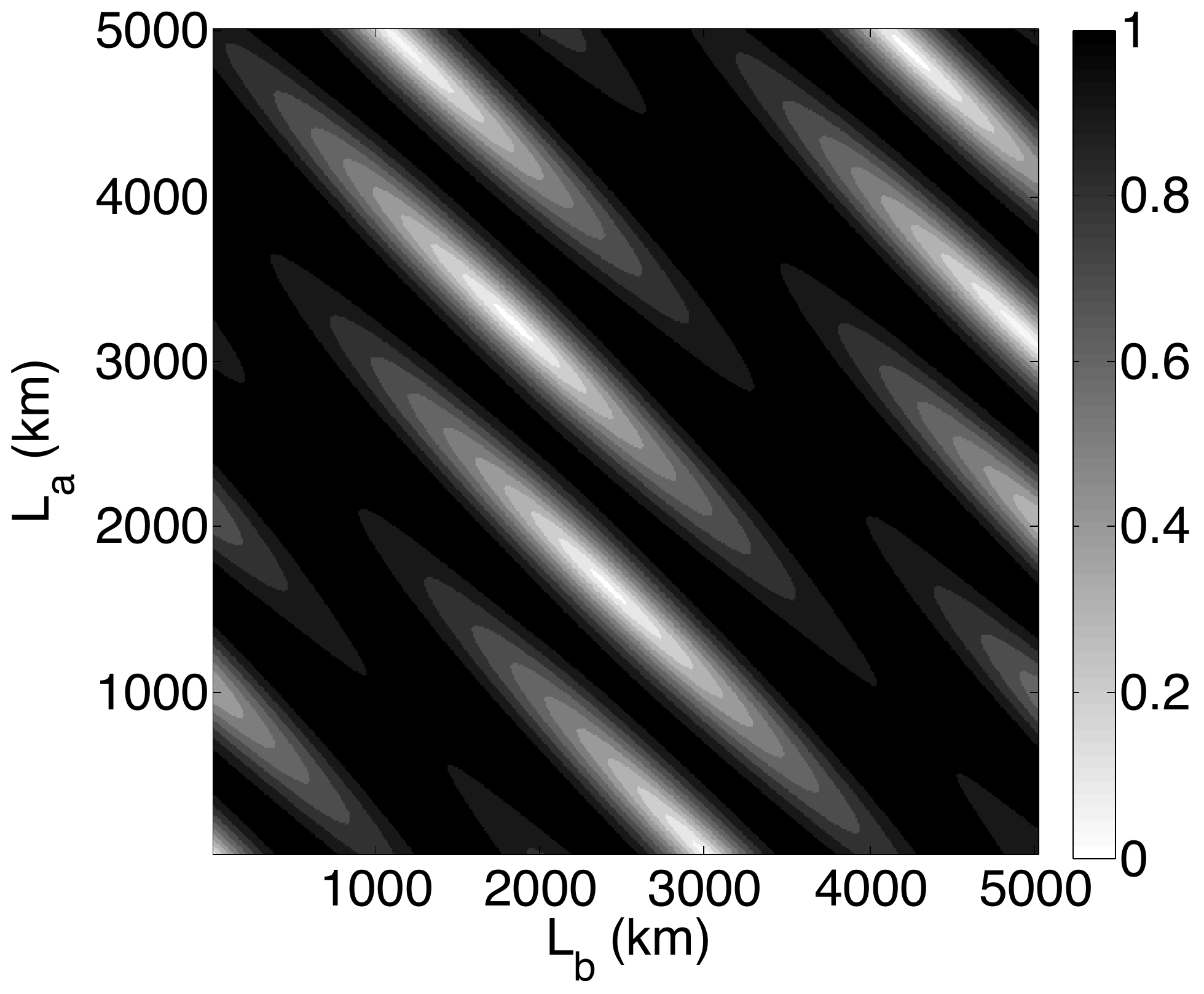}
\caption{ A plot of $\overline u$ for $E=1$ GeV and various baselines $L_{a,b}$ with $\theta=0.59$, $\phi=0.15$, and $\psi = 0.72$ in the castle-wall profile. 
 \label{fig11}}
\end{figure}

\begin{figure}
\includegraphics[width=8.6cm]{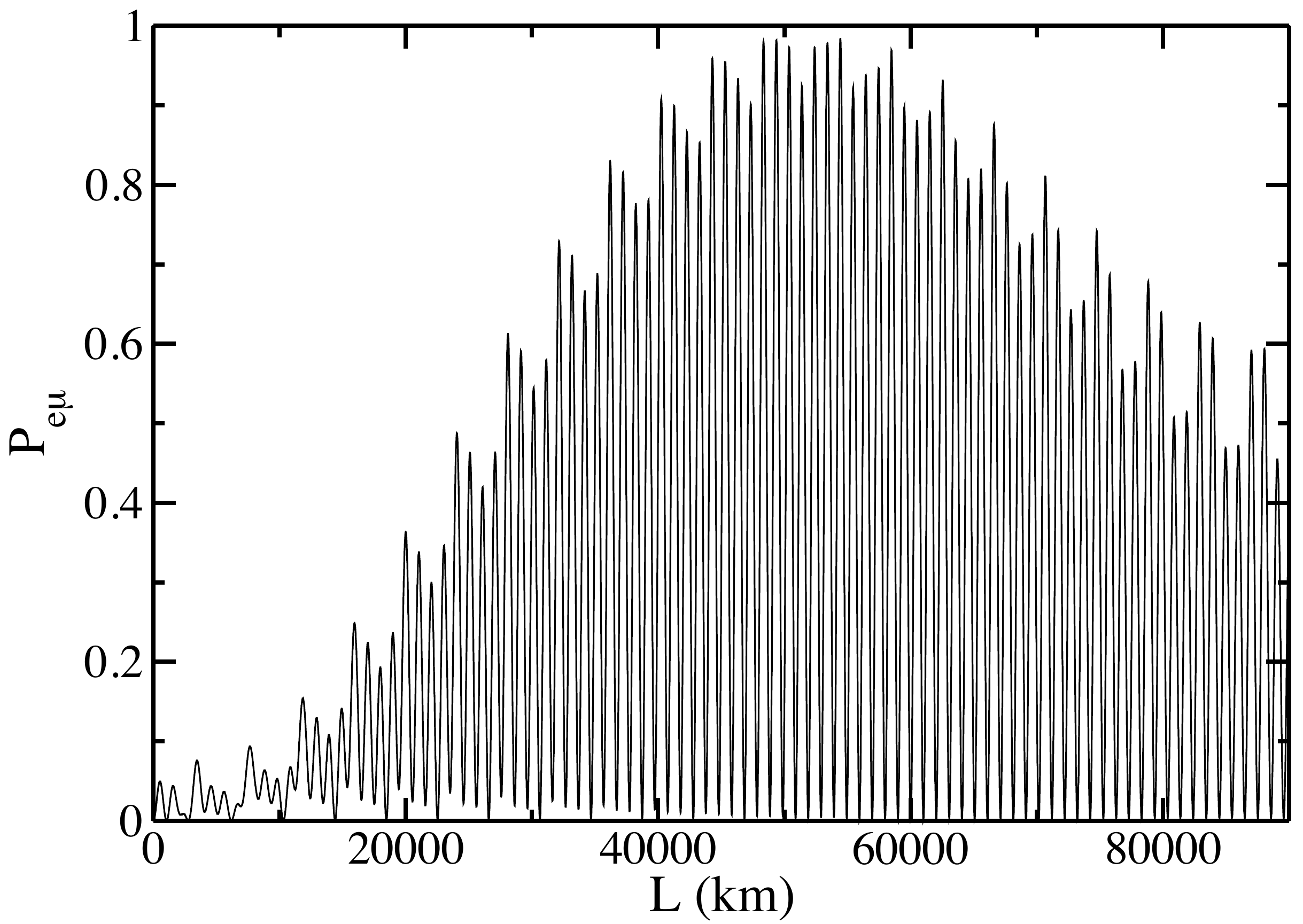}
\caption{ The oscillation probability $\mathcal{P}_{e\mu}$ for 1 GeV neutrinos through a castle-wall profile with $L_a = 1625$ km and $L_b = 2410$ km.  We set $\theta = 0.59$, $\phi = 0.15$, and $\psi = 0.72$. We omit the shaded regions which typically indicate the higher density region. \label{fig12}}
\end{figure}

More generally, with non-zero $\phi$, we can adapt the results from Case (ii) to determine a parametric resonance condition for $\hat \nu$.   For the given choice of parameters, the parametric resonance condition does not change significantly. The points at which $\overline u$ attains a minimum in $L_a$-$L_b$ parameter still correspond to solutions of Eqs.~(\ref{2nu_prc},\ref{u4_iii}), but the region in which $\overline u \le 0.1$ shrinks considerably.  We plot $\overline u$ for 1 GeV neutrinos for the castle-wall profile with $\theta=0.59$, $\phi=0.15$, and $\psi = 0.72$, Fig.~\ref{fig11}.
With all three mixing angles non-zero, genuine three-neutrino oscillations is present in all channels.  In particular, for $\nu_e \to \nu_\mu$ oscillations two oscillation scales are now present.  With non-zero $\phi$, this oscillation probability can approach unity.  As an example, we choose the baselines $L_a = 1625$ km and $L_b = 2410$ km which correspond to a local minimum of $\overline u$ in Fig.~\ref{fig11}.  We plot in Fig.~\ref{fig12} the oscillation probability $\mathcal{P}_{e \mu}$ for this castle-wall profile.  The oscillation probability attains a maximum value of 0.98.  With the extra oscillation channel, we are able to evade the limit set by $\cos^2 \psi$.

\section{Core-crossing trajectories}

In the previous sections, the neutrino oscillation probability could increase dramatically (relative to constant-density trajectories) after propagating through several periods of a castle-wall potential.  For neutrinos with energies on the  order of a few hundred MeV, the number of periods needed to attain a maximum value of $\mathcal{P}_{e\mu}$ is small since the effective matter mixing angle $\tilde \theta_{12}$ is relatively large.  For a neutrino energy around 1 GeV, the effective mixing angle decreases dramatically, and the number of periods needed to achieve the maximum oscillation probability increases in turn.  The reality is that the baselines for {\em one} period are on the order of the Earth's diameter, rendering multiple baselines impossible in a laboratory setting.  At best, the Earth's density profile can be used as a laboratory.  We can approximate the Earth's interior as a high density core ($\rho_b \sim 11.5$ g/cm$^3$) of radius  $R_c =3485$ km surrounded by a mantle of density $\rho_a = 4.5$ g/cm$^3$ \cite{prem}.  For a detector located near the surface of the Earth, an atmospheric neutrino passing through the Earth to the detector will travel along a chord which can be parametrized by the zenith angle $\Theta$.  Upgoing neutrinos, which travel the Earth's diameter,  have $\Theta=\pi$ and thus $\cos \Theta = -1$. For this trajectory,  the initial baseline through the mantle is $L_a = R_e - R_c = 2886$ km where the radius of the earth is $R_e = 6371$ km, and then the path through the core is given by its diameter $L_b = 2 R_c =6970$ km.  As the zenith angle decreases, the distance traveled through the mantle increases, while the distance through the core decreases; generally, we have
\begin{eqnarray}
 L_a&=&-R_e \cos\Theta-\sqrt {R_c^2-(R_e \sin\Theta)^2},	\\
  L_b&=&2\sqrt {R_c^2-(R_e\sin\Theta)^2}.
\end{eqnarray}
For zenith angles less than $147^\circ$ (or $\cos \Theta > -0.84$), the neutrino does not travel through the core. 

\begin{figure}
\includegraphics[width=8.6cm]{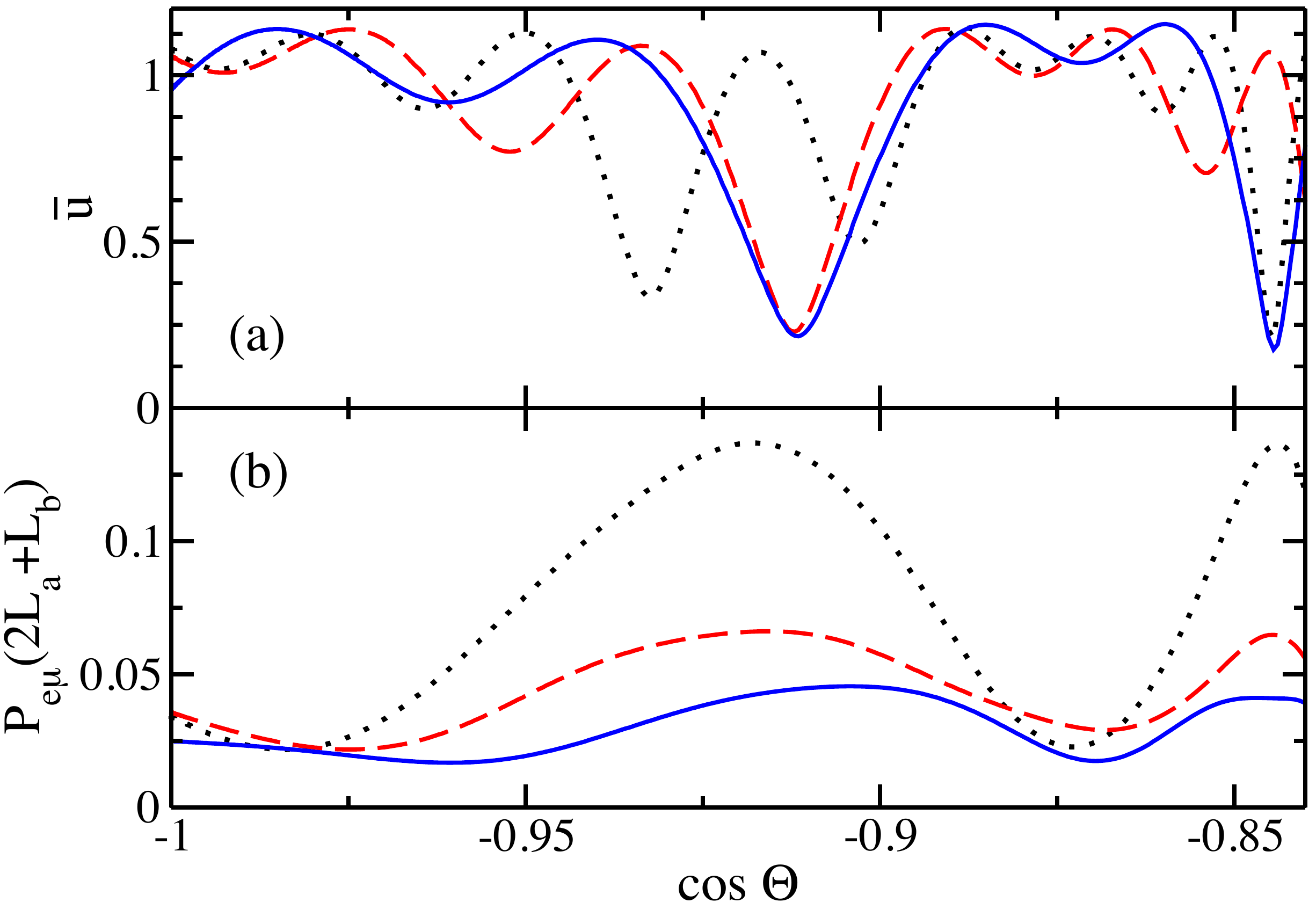}
\caption{ (Color online) (a) We plot $\overline{u}$ for core-crossing trajectories.  The dotted (black) curve is for neutrino energy 600 MeV; the dashed (red) curve is for neutrino energy 800 MeV; the solid (blue) curve is for neutrino energy 1 GeV.  (b) The oscillation probability $\mathcal{P}_{e\mu}$ at a detector after traveling a distance of $2L_a+L_b$ along a core-crossing trajectory.  The curves are averaged over a flat energy spectrum with a width of 200 MeV.
The dotted (black) curve's spectrum is centered on 600 MeV; the dashed (red) curve's spectrum is centered on 800 MeV; the solid (blue) curve's spectrum is centered on 1 GeV.    \label{fig13}
}
\end{figure}

For core-crossing trajectories, we compute $\overline u$ as a function of the zenith angle $\Theta$ for energies between 600 MeV and 1 GeV, Fig.~\ref{fig13}(a).  This parameter, $\overline u$, does attain local minima for these energies along chords with a zenith angle that satisfies $-0.95 \le \cos \Theta \le -0.9$.  The absolute minima do not vanish, yet they are sufficiently small to result in some parametric enhancement.  This enhancement is apparent when considering the oscillation probability $\mathcal{P}_{e\mu}$ at the terminus of the trajectory, Fig.~\ref{fig13}(b).  Here, we consider the value of the oscillation probability averaged over a flat energy spectrum with width 200 MeV centered on the same energies considered in Fig.~\ref{fig13}(a).  The overall scale of the oscillation probability is significantly suppressed since the energies under consideration are well beyond the MSW resonance where the mixing angle decreases inversely with the energy, $\tilde \theta_{12} \sim \theta_{12} \Delta_{21}/(2EV)$.  Regardless, parametric effects significantly enhance the probability relative to neutrinos traveling through a constant density mantle along the same baselines.  For neutrino beams centered around 600 MeV, 800 Mev, or 1 GeV the average oscillation probability $\mathcal{P}_{e\mu}$ through a constant density mantle would yield maximal values of 0.05, 0.03, and 0.02  (respectively).  Through the core-crossing trajectory, in the region of parametric resonance, the oscillation probability is enhanced by a factor of two to three.  While the absolute value is small, in a high precision experiment involving upgoing atmospheric neutrinos the effect can be relevant.

\section{Conclusion}

We have examined parametric resonance in a full three-neutrino framework.  To do so, we found it necessary to simplify the expression for the time-evolution operator  in constant density matter  for the three-neutrino state.   With this simpler expression, we considered the castle-wall matter profile and determined, for given values of the oscillation parameters and profile densities, what appropriate baselines would lead to parametric resonance.  We focused on neutrino energies in an intermediate range from a few hundred MeV to a few GeV; i.e., the term $2EV$ is large relative to $\Delta_{21}$ but small relative to $\Delta_{31}$.  Since $\theta_{13}$ is small, we are able to consider its effects perturbatively.  We found that the parametric resonance condition was a confluence of two conditions related to the two different oscillation scales.  The two-neutrino condition essentially carries over to the three-neutrino framework; however, when $\theta_{13}$ is nonzero, the other oscillation scale must be considered.  If the new parametric resonance condition is identically satisfied, the oscillation probability $\mathcal{P}_{e\mu}$ tends to unity after the neutrinos travel through a number of periods of the matter profile.  If the condition is just approximately satisfied, full parametric resonance is not achieved; however, the oscillation probability is still enhanced in the periodic matter profile, relative to a trajectory through constant density matter.  This is the situation for sub-GeV atmospheric neutrinos which travel through the Earth's core.  For such core-crossing trajectories, the parameter $\overline u$ attains a minimum value on the order of 0.2.  Despite this, the oscillation probability for these trajectories is significantly enhanced.
\bibliography{biblio}

\end{document}